\documentclass[a4paper,12pt]{article}

\usepackage[a4paper,text={16.8cm,22.4cm}]{geometry}
\usepackage{amsmath,amssymb,bm,graphicx,color}
\usepackage{extarrows}

\allowdisplaybreaks 
\newcommand{\eps}{\varepsilon}
\newcommand{\taubar}{\bar\tau}
\newcommand{\altsqrt}[1]{\sqrt{\rule{0mm}{3.5mm} #1}}
\def\rd{\mathrm{d}}

\begin{document}

\begin{titlepage}

\vspace*{-1.0cm}
\begin{flushright}
SI-HEP-2018-16 \\
QFET-2018-10 \\ 
DESY 18-078
\end{flushright}

\vspace{1.2cm}
\begin{center}
\Large\bf
\boldmath
Two-loop anomalous dimensions of \\
generic dijet soft functions
\unboldmath
\end{center}

\vspace{0.2cm}
\begin{center}
Guido Bell$^a$, Rudi Rahn$^b$ and Jim Talbert$^c$\\
\vspace{0.4cm}
{\sl 
${}^a$\,Theoretische Physik 1, Naturwissenschaftlich-Technische Fakult\"at,\\
Universit\"at Siegen, Walter-Flex-Strasse 3, 57068 Siegen, Germany\\[0.3cm]
${}^b$\,Albert Einstein Center for Fundamental Physics, 
Institut f\"ur Theoretische Physik, Universit\"at Bern,
Sidlerstrasse 5, 3012 Bern, Switzerland\\[0.3cm]
${}^c$\,Theory Group, Deutsches Elektronen-Synchrotron (DESY), 22607 Hamburg, Germany}
\end{center}

\vspace{0.5cm}
\begin{abstract}
\vspace{0.2cm}
\noindent 
We present compact integral representations for the calculation of two-loop anomalous 
dimensions for a generic class of soft functions that are defined in terms of two 
light-like Wilson lines. Our results are relevant for the resummation of Sudakov 
logarithms for $e^+ e^-$ event-shape variables and inclusive hadron-collider observables
at next-to-next-to-leading logarithmic accuracy within Soft-Collinear Effective 
Theory (SCET). Our formalism applies to both SCET-1 and SCET-2 soft functions and we 
clarify the relation between the respective soft anomalous dimension and the collinear 
anomaly exponent. We confirm existing two-loop results for about a dozen dijet soft 
functions and obtain new predictions for the angularity event shape and the
soft-drop jet-grooming algorithm.
\end{abstract}
\vfil

\end{titlepage}

\section{Dijet soft functions}

Scattering cross sections at large momentum transfer $Q$ are often sensitive to large 
logarithmic corrections that spoil the convergence of the perturbative expansion in 
the strong coupling $\alpha_s(Q) \ll 1$. By computing corrections of the form 
$\alpha_s(Q) L\sim 1$ to all orders, where $L\gg1$ represents the large logarithm, 
the theoretical predictions can be systematically improved with respect to a fixed-order 
expansion. This re\-organisation of the perturbative series -- commonly called 
\emph{resummation} --  can be achieved on the basis of factorisation theorems which
disentangle the relevant scales of the scattering process to all orders in perturbation 
theory. 

The factorisation of cross sections in QCD has a long history. Traditionally, factorisation 
was established via an analysis of Feynman diagrams that incorporates the constraints from 
gauge invariance using Ward identities (see~\cite{Collins:1989gx,Collins:2011zzd} for a 
review). Alternatively, the problem can be accessed with methods from effective field theory, 
which separate the effects from the relevant degrees of freedom directly on the level of the 
Lagrangian. The two approaches have many similarities and yield identical physical results 
(see e.g.~\cite{Almeida:2014uva} for a detailed comparison). While we use the language of 
Soft-Collinear Effective Theory (SCET)~\cite{Bauer:2000yr,Bauer:2001yt,Beneke:2002ph} in the 
present work, we stress that our analysis is also relevant for resummations that are formulated 
in QCD. 

The scattering processes of interest in this work involve two hard, massless and colour-charged 
partons at the Born level. Whenever the QCD radiation is confined to be low-energetic (soft) 
or collinear to the directions of the hard partons, the partonic cross section factorises 
in the schematic form
\begin{align}
d\hat\sigma = H \cdot J_n \otimes J_{\bar n} \otimes S\,,
\label{eq:fact}
\end{align}
where the symbol $\otimes$ denotes a convolution in suitable kinematic variables. The hard 
function $H$ contains the virtual corrections to the Born process at the scale $Q^2$, the 
jet functions $J_n$ and $J_{\bar n}$ encode the effects from the collinear emissions in 
the directions $n^\mu$ and $\bar n^\mu$ of the hard partons, and the soft function $S$ 
describes the low-energetic cross talk between the two jets. The characteristic scales 
associated with the jet and soft functions are typically much smaller than $Q^2$, but their 
respective hierarchy depends on the specific observable. In fact, different 
hierarchies between the jet and soft scales are described by different versions of the 
effective theory as we will see below.

The individual factors in \eqref{eq:fact} depend on an unphysical factorisation scale,
and by solving the associated renormalisation group equations (RGEs) one can resum the 
logarithmic corrections to the cross section to all orders. Whereas the fixed-order 
expansion is organised into leading order (LO) corrections, next-to-leading order (NLO) 
corrections, and so on, the resummed expressions refer to leading logarithmic (LL) accuracy, 
next-to-leading logarithmic (NLL) accuracy, etc. As in any effective field theory, the 
desired accuracy can then be achieved by computing anomalous dimensions and matching 
corrections to a given order in perturbation theory. For Sudakov problems with a double 
logarithm per loop order, the appropriate counting scheme is given e.g.~in Table~5 
of~\cite{Almeida:2014uva}.

The purpose of this work is to develop a systematic framework for the computation
of two-loop soft anomalous dimensions for a wide class of collider observables. The two-loop 
soft anomalous dimension is required for NNLL resummation and it often represents the only 
missing piece at this accuracy, since the two-loop hard anomalous dimension is known 
for arbitrary processes~\cite{Becher:2009qa} and the two-loop jet anomalous dimension can 
then be extracted from the factorisation theorem \eqref{eq:fact} using RG invariance of the 
cross section. While one in addition needs the one-loop hard, jet, and soft matching corrections 
at NNLL, their computation often represents a comparably simple task.

The soft functions that enter the factorisation theorem \eqref{eq:fact} are given by vacuum 
matrix elements of a configuration of Wilson lines that reflect the structure of the 
scattering process at the Born level. More specifically, they can be written in the form
\begin{align}
S(\tau, \mu) = \frac{1}{N_c} \; \sum_X \; 
\mathcal{M}(\tau;\lbrace k_{i} \rbrace)\;
\mathrm{Tr}\; 
|\langle X | \,T [S^{\dagger}_{n}(0) S_{\bar{n}}(0)]\, | 0 \rangle |^{2}\,,
\label{eq:softfun:definition}
\end{align}
where $S_{n}$ and $S_{\bar{n}}$ are soft Wilson lines extending along two light-like 
directions $n^{\mu}$ and $\bar{n}^{\mu}$ with $n\cdot{\bar{n}}=2$. For concreteness, 
we assume that the Wilson lines are in the fundamental colour representation and the 
definition in \eqref{eq:softfun:definition} involves  a trace over colour indices 
as well as a generic measurement function $\mathcal{M}(\tau,\lbrace k_{i} \rbrace)$ 
that provides a constraint on the soft radiation with parton momenta 
$\lbrace k_{i} \rbrace$ according to the observable under consideration. The explicit 
form we assume for the measurement function will be specified in the following section. 
Notice that up to two-loop order, it is irrelevant whether $n^{\mu}$ and $\bar{n}^{\mu}$ 
refer to incoming or outgoing directions~\cite{Catani:2000pi,Kang:2015moa}, and our results 
therefore equally apply to $e^+e^-$ dijet observables, one-jet observables in 
deep-inelastic scattering or zero-jet observables at hadron colliders. For convenience, 
we refer to all of these cases as \emph{dijet soft functions} in the following.

The soft functions can further be classified according to the hierarchy between the jet
and soft scales in the under\-lying factorisation theorem \eqref{eq:fact}. 
For SCET-1 observables, the virtuality 
of the collinear modes is much larger than the one of the soft modes and the logarithmic 
corrections can be resummed using standard RG techniques. For SCET-2 observables, 
on the other hand, the jet and soft scales are of the same order and additional techniques 
like the collinear anomaly~\cite{Becher:2010tm} or the rapidity RG~\cite{Chiu:2012ir} 
are needed to resum the logarithmic corrections. On the technical level, the difference 
between SCET-1 and SCET-2 soft functions manifests itself in the form of rapidity 
divergences in the phase-space integrals that are not regularised in dimensional 
regularisation. One therefore needs to introduce an additional regulator for SCET-2 soft 
functions, for which we use a symmetrised version of the analytic regulator proposed 
in~\cite{Becher:2011dz}.

The outline of this paper is as follows: In the next section we define more precisely 
which class of dijet soft functions we consider by specifying the functional form we 
assume for the measurement function. In Section~\ref{sec:SCET-1} we present our results for 
the calculation of two-loop soft anomalous dimensions for SCET-1 type observables, and 
Section~\ref{sec:SCET-2} contains the corresponding expressions for SCET-2 soft functions, 
for which the relevant anomalous dimension is often called the collinear anomaly exponent. 
The expressions we find in the SCET-1 and SCET-2 sections turn out to be similar, and we 
elaborate on the relation between the soft anomalous dimension and the collinear anomaly 
exponent in Section~\ref{sec:SCET-1vsSCET-2}. In Section~\ref{sec:generalisation} we discuss 
several extensions of our formalism which are relevant, e.g.,~for jet-veto observables, 
multi-differential soft functions, and processes with more than two jet directions. 
We finally conclude in Section~\ref{sec:conclusions} and present further details of our 
calculation in two appendices.

\section{Measurement function}
\label{sec:measurement}

The soft functions we consider are typically defined in Laplace (or Fourier) space. 
The main reason why we work in Laplace space is that the factorisation theorem \eqref{eq:fact} 
and the associated RGEs often take a particularly simple form in this space. For some soft 
functions like those associated with jet-veto observables, the RGEs are more naturally 
formulated in momentum (or cumulant) space, but even in this case it is possible to work 
with the Laplace transform to bring the soft function into the form considered in this 
section, and to correct for the factors associated with the inversion of the Laplace 
transformation at a later stage. We will come back to the discussion of jet-veto 
observables in Section~\ref{sec:generalisation}.

Another advantage of the Laplace space technique is that the functions are not 
distribution-valued. At tree level the measurement function is then trivial and can be 
normalised to one, $\mathcal{M}_0(\tau)=1$. For a single emission with momentum $k^\mu$, 
we introduce light-cone coordinates with $k_+ = n \cdot k$, $k_- = \bar n \cdot k$ and a 
vector $k_\perp^\mu$ that is transverse to $n^\mu$ and $\bar n^\mu$. We further parametrise 
the phase-space integrals in terms of the  magnitude of the transverse momentum $k_T$,
a measure of the rapidity $y_k$, and an angular variable $t_k$ as
\begin{align}
\label{eq:param:NLO}
k_{-} = \frac{k_{T}}{\sqrt{y_k}}\,, \qquad\qquad 
k_{+} = \sqrt{y_k} \,k_{T}\,, \qquad\qquad
\cos\theta_k = 1-2t_k\,.
\end{align}
A non-trivial dependence on the angle $\theta_k$ may arise when the measurement is performed 
with respect 
to a vector $v^\mu$ that differs from the jet axes $n^\mu$ and $\bar n^\mu$. If so, we 
project this vector onto the transverse plane and denote the angle between 
$\vec{v}_\perp$ and $\vec{k}_\perp$ by $\theta_k$.

In terms of these variables, we assume that the single-emission measurement function can 
be written in the form
\begin{align}
\label{eq:measure:NLO}
\mathcal{M}_1(\tau; k) = \exp\big(-\tau\, k_{T}\, y_k^{n/2}\, f(y_k,t_k)\,\big)\,,
\end{align}
where the exponential reflects the fact the we work in Laplace space. We further assume 
that the Laplace variable $\tau$ has the dimension $1/$mass, which fixes the linear 
dependence on the variable $k_T$ on dimensional grounds. At NLO
the soft functions we consider are thus characterised by a parameter $n$ and a function 
$f(y_k,t_k)$ that encodes the angular and rapidity dependence\footnote{We assume that the 
real part of the function $f(y_k,t_k)$ is positive, since the $k_T$ integral would 
otherwise not converge. The same assumption applies to the functions $F$ and $G$ in 
equations (\ref{eq:measure:NNLO:corr}) and (\ref{eq:measure:NNLO:unc}) below. The functions 
$f$, $F$ and $G$ should furthermore be independent of the dimensional and the rapidity 
regulators.}. One can show that the parameter $n$ is related to the power counting of the 
soft modes in the underlying factorisation theorem and that the value $n=0$ corresponds to 
a SCET-2 observable~\cite{Bell:2018oqa}. 

For our purposes, it is sufficient to adopt a pragmatic approach to determine the 
parameter $n$ for a given observable. After integration over $k_T$ and expanding in the 
various regulators, the expression contains logarithms of the function $f(y_k,t_k)$ that 
are multiplied by a matrix element that is divergent in the collinear limit $y_k\to0$. It 
is therefore crucial to factor out the leading scaling in $y_k$, i.e.~we \emph{define} the 
parameter $n$ by the requirement that the function $f(y_k,t_k)$ is finite and non-zero in 
the limit $y_k\to0$. 

The considered class of soft functions may look specific, but it captures a large variety 
of dijet soft functions as will become clear when we discuss explicit examples below. 
Sample expressions for the parameter $n$ and the associated function $f(y_k,t_k)$ for various 
$e^+e^-$ and hadron-collider soft functions can be found in Table~1 of~\cite{Bell:2015lsf}.

At NNLO one in addition needs to specify the double-emission measurement function. As the 
singularity structure of the underlying matrix element differs among the colour structures, 
we apply distinct phase-space parametrisations for the correlated ($C_F T_F n_f$, $C_F C_A$) 
and uncorrelated ($C_F^2$) emission contributions. These parametrisations have been chosen 
according to two criteria: First, they should allow us to factorise the divergences of the 
matrix elements and, second, they should provide a simple parametrisation of the measurement 
function with a two-emission equivalent of the function $f(y_k,t_k)$ that is finite in the 
singular limits of the matrix element. We found it further\-more convenient to exploit the 
symmetries from $n\leftrightarrow\bar n$ and $k\leftrightarrow l$ exchange, where $k$ and 
$l$ are the momenta of the emitted partons, to map the integration region onto the unit 
hypercube~\cite{Bell:2018oqa}.

For the correlated double-emission contribution, we parametrise
\begin{align}
\label{eq:param:NNLO:corr}
k_- = \frac{a b}{1+a b} \;\frac{p_{T}}{\sqrt{y}}\,, \qquad
k_+ = \frac{b}{a+b} \;\sqrt{y}\,p_{T}  \,, \qquad
l_- = \frac{1}{1+a b} \;\frac{p_{T}}{\sqrt{y}}\,, \qquad
l_+ = \frac{a}{a+b} \;\sqrt{y}\,p_{T}  \,,
\end{align}
where $p_T$ and $y$ are functions of the sum of the light-cone momenta, $a$ is a measure of 
the rapidity difference of the emitted partons, and $b$ is the ratio of their transverse 
momenta~\cite{Bell:2015lsf}\footnote{The variable $p_T=\sqrt{(k_-+l_-)(k_++l_+)}$ 
should not be confused with the total transverse momentum of the emitted partons.}. 
In general the measurement function now depends on three angles
$\theta_k=\sphericalangle(\vec{v}_\perp,\vec{k}_\perp)$,
\mbox{$\theta_l=\sphericalangle(\vec{v}_\perp,\vec{l}_\perp)$}
and $\theta_{kl}=\sphericalangle(\vec{k}_\perp,\vec{l}_\perp)$, and we denote the 
corresponding variables that are defined on the unit hyper\-cube by $t_k$, $t_l$ and $t_{kl}$
in analogy to \eqref{eq:param:NLO}. The corresponding relation to (\ref{eq:measure:NLO}) 
for the correlated double-emission contribution then becomes
\begin{align}
\label{eq:measure:NNLO:corr}
\mathcal{M}_2^{corr}(\tau; k,l) = \exp\big(-\tau\, p_{T}\, y^{n/2}\, 
F(a,b,y,t_k,t_l,t_{kl})\,\big)\,,
\end{align}
where the dependence on $p_T$ is again fixed on dimensional grounds and the function 
$F$ is assumed to be finite and non-zero in the limit $y\to0$. Notice that 
this is achieved by factorising the same power of the rapidity variable $y$ as in the 
one-emission case~\cite{Bell:2018oqa}.

For uncorrelated emissions we use a phase-space parametrisation that itself depends on the 
parameter $n$,
\begin{align}
\label{eq:param:NNLO:unc}
k_- &= \left( \frac{\sqrt{y_l}}{1+y_l} \right)^{n}
\frac{b}{1+b} \;\frac{q_{T}}{\sqrt{y_k}}\,, \hspace{2.1cm}
l_- = \left( \frac{\sqrt{y_k}}{1+y_k} \right)^{n}
\frac{1}{1+b} \;\frac{q_{T}}{\sqrt{y_l}}\,,
\nonumber\\
k_+ &= \left( \frac{\sqrt{y_l}}{1+y_l} \right)^{n}
\frac{b}{1+b} \;\sqrt{y_k}\,q_{T}\,, \qquad\qquad
l_+ = \left( \frac{\sqrt{y_k}}{1+y_k} \right)^{n}
\frac{1}{1+b} \;\sqrt{y_l}\,q_{T}\,,
\end{align}
where $q_T$ is now the only dimensionful variable, $y_k$ and $y_l$ are measures of the 
rapidities of the individual partons, and $b$ reduces to the ratio of their transverse 
momenta for $n=0$ (the parentheses introduce rapidity-dependent weight factors for 
$n\neq0$)~\cite{Bell:2018jvf}. 
The measurement function for uncorrelated emissions is then parametrised as
\begin{align}
\label{eq:measure:NNLO:unc}
\mathcal{M}_2^{unc}(\tau; k,l) = \exp\big(-\tau\, q_{T}\, y_k^{n/2}\, y_l^{n/2}\, 
G(y_k,y_l,b,t_k,t_l,t_{kl})\,\big)\,,
\end{align}
where the dependence on $q_T$ is once more fixed on dimensional grounds and the function 
$G$ is supposed to be finite and non-zero in the collinear limits $y_k\to0$ and $y_l\to0$.
The latter again requires us to factorise the rapidity variables $y_k$ and $y_l$
to the same power as in \eqref{eq:measure:NLO}.

Up to NNLO the considered class of soft functions is thus characterised by the three 
functions $f(y_k,t_k)$, $F(a,b,y,t_k,t_l,t_{kl})$, $G(y_k,y_l,b,t_k,t_l,t_{kl})$ and 
a parameter $n$. As an example, we consider the soft function for $W$-production at 
large transverse momentum discussed in~\cite{Becher:2012za}\footnote{This example is 
strictly speaking not a dijet soft function since the definition involves Wilson lines 
in three light-like directions, namely two beam directions $n_1$ and $n_2$ and the direction
of a jet $n_J$ that recoils against the $W$-boson. It has been shown, however, 
in~\cite{Becher:2012za} that the gluon attachments to the Wilson line $S_{n_J}$ vanish up 
to NNLO and one is furthermore free to choose $n_1\cdot n_2=2$ along with 
$n_1\cdot n_J=n_2\cdot n_J=2$ due to rescaling invariance of the Wilson lines. The soft 
function is therefore of the dijet-type considered here and the vector $n_J$ introduces a 
non-trivial angular dependence.}. In Laplace space the one-emission measurement function 
reads $\mathcal{M}_1(\tau; k) = \exp\big(-\tau\, n_J \cdot k\big)$, from which we read 
off that the jet direction $n_J^\mu$ serves as the measurement vector $v^\mu$ in this 
case. After decomposing $n_J^\mu$ in light-cone coordinates, one has 
$n_J \cdot k = k_- + k_+ - 2 k_T \cos \theta_k$, which in the parametrisation 
(\ref{eq:param:NLO}) leads to $n=-1$ and $f(y_k,t_k) = 1 + y_k - 2\sqrt{y_k}(1-2t_k)$. 
For two emissions, the measurement function involves the sum $n_J \cdot k + n_J \cdot l$, 
which for correlated emissions implies
\begin{align}
F(a,b,y,t_k,t_l,t_{kl}) = 
1 + y - 2 \;\sqrt{\frac{a y}{(1+a b)(a+b)}} \;
\Big(b (1-2t_k) + 1-2t_l\Big)\,,
\end{align}
which is finite in the limit $y\to0$. For uncorrelated emissions, one obtains 
\begin{align}
G(y_k,y_l,b,t_k,t_l,t_{kl})=
\frac{b(1+y_l)(1 + y_k - 2\sqrt{y_k}(1-2t_k))}{(1+b)} 
+\frac{(1+y_k)(1 + y_l - 2\sqrt{y_l}(1-2t_l))}{(1+b)} \,,
\end{align}
which is again finite in the limits $y_k\to0$ and $y_l\to0$.

In general the functions $F$ and $G$ are constrained by infra\-red and collinear safety. 
In the soft limit $k^\mu\to0$, which corresponds to the limit $b\to0$ in our 
parametrisations, one has
\begin{align}
F(a,0,y,t_k,t_l,t_{kl}) = f(y,t_l)\,,
\qquad\qquad
G(y_k,y_l,0,t_k,t_l,t_{kl}) = \frac{f(y_l,t_l)}{(1+y_k)^n}\,.
\end{align}
After using the $k\leftrightarrow l$ symmetry, the soft limit $l^\mu\to0$ is mapped onto the 
same constraints. Whenever the two emitted partons become collinear to each other, one obtains 
\begin{align}
F(1,b,y,t_l,t_l,0) = f(y,t_l)\,,\qquad\qquad
G(y_l,y_l,b,t_l,t_l,0) = \frac{f(y_l,t_l)}{(1+y_l)^n}\,.
\end{align}
We use these relations in the following to verify if the poles of the bare soft function 
cancel as predicted by the RGE. The constraints also serve as a check for the derivation 
of the functions $F$ and $G$, and one easily verifies that they are satisfied for the 
example from above.\\[0.8em]

With the phase-space parametrisations and the measurement function at hand, the soft 
function can be evaluated in dimensional regularisation with $d=4-2\varepsilon$ dimensions. 
The basic strategy for the evaluation of the integrals has been outlined 
in~\cite{Bell:2015lsf, Bell:2018jvf} and further details will be given in a future 
publication~\cite{Bell:2018oqa}. For SCET-2 soft functions with $n=0$, we implement a variant of 
the phase-space regulator proposed in~\cite{Becher:2011dz},
\begin{equation}
\int\!d^dp \; \left(\frac{\nu}{p_++p_-}\right)^\alpha \;  \delta(p^2) \theta(p^0) \,,
\end{equation}
which respects the $n\leftrightarrow\bar n$ symmetry. The rapidity divergences then manifest 
themselves as poles in the regulator $\alpha$.

\section{Soft anomalous dimension}
\label{sec:SCET-1}

For observables with $n\neq0$, the phase-space integrals are well defined in dimensional 
regularisation and the soft function is defined in SCET-1. Our goal then consists in 
determining the soft anomalous dimension $\gamma^{S}(\alpha_s)$ from the $1/\eps$ poles 
of the bare soft function. In the following we assume that the soft function renormalises 
multiplicatively in Laplace space, and that the RGE can be written in the form
\begin{align}
\label{eq:RGE:SCET-1}
\frac{\rd}{\rd \ln\mu}  \; S(\tau,\mu)
&= - \frac{1}{n} \,\bigg[ 4 \,\Gamma_{\mathrm{cusp}}(\alpha_s) \, 
\ln(\mu\taubar) 
-2 \gamma^{S}(\alpha_s) \bigg] \; S(\tau,\mu) \,,
\end{align}
where $\taubar = \tau e^{\gamma_E}$ and $\Gamma_{\mathrm{cusp}}(\alpha_s)$ is the universal 
cusp anomalous dimension. Notice that we define the soft anomalous dimension with a prefactor 
$2/n$, where $n$ reflects the scaling of the observable in the soft-collinear limit as 
discussed in the previous section. The leading coefficients in the expansion of the cusp 
anomalous dimension 
$\Gamma_{\mathrm{cusp}}(\alpha_s)= \sum_{n=0}^\infty \,\Gamma_n (\frac{\alpha_s}{4\pi})^{n+1}$ 
are $\Gamma_0 = 4 C_F$ and
$\Gamma_1/\Gamma_0 = (67/9 - \pi^2/3 ) C_A - 20/9\, T_F n_f.$

Expanding the soft anomalous dimension as 
$\gamma^{S}(\alpha_s) = \sum_{n=0}^\infty \,\gamma^{S}_n \,(\frac{\alpha_s}{4\pi})^{n+1}$, 
we can determine its leading coefficient from the NLO calculation that has been described in 
detail in~\cite{Bell:2015lsf}. For the class of soft functions defined in 
(\ref{eq:measure:NLO}), we find
\begin{align} 
\gamma^{S}_0 &= -\frac{16C_F}{\pi}
\int_0^1 \!dt_k \;\,
\frac{\ln f(0,t_k)}{\altsqrt{4t_k\bar t_k}}
\label{eq:gamma0}
\end{align}
with $\bar t_k=1-t_k$. Notice that the single-emission function $f(y_k,t_k)$ enters this 
formula only in the collinear limit $y_k\to0$, which explains why the one-loop anomalous 
dimension is identical for many observables. With the normalisation adopted in 
(\ref{eq:RGE:SCET-1}), $\gamma^{S}_0$ is moreover independent of the parameter $n$.

At NNLO the soft anomalous dimension receives contributions from three colour structures,
\begin{align} 
\label{eq:gamma1:colour}
\gamma_1^S &= \gamma_1^{n_f} \,C_F T_F n_f
\,+\, \gamma_1^{C_A} \,C_F C_A 
\,+\,  \gamma_1^{C_F} \,C_F^2\,.
\end{align}
The first two terms refer to the correlated emission contribution, which according to 
(\ref{eq:measure:NNLO:corr}) can be expressed in terms of the function 
$F(a,b,y,t_k,t_l,t_{kl})$. Similar to the one-loop result, it turns out that this function 
is only required in the limit $y\to0$, and we obtain
\begin{align}
\gamma^{n_f}_1 &= \frac{224}{27} - \frac{4\pi^2}{9} 
 + \frac{64}{9\pi}\,
\int_0^1 \!dt_l \;\,
\frac{5+3\ln(16t_l\bar t_l)}{\altsqrt{4t_l\bar t_l}}\;
\ln f(0,t_l) 
\nonumber\\[0.2em]
&\quad
+\frac{1}{\pi^2}\int_0^1 \!da \int_0^1 \!db \int_0^1 \!dt_l \int_0^1 \!dt_{kl} \;
\frac{k_{1}(a,b,t_{kl})}{\altsqrt{16t_l\bar t_l t_{kl}\bar t_{kl}}}\; \,
\mathcal{F}(a,b,t_l,t_{kl})\,,
\nonumber\\[0.2em]
\gamma^{C_A}_1 &= -\frac{808}{27} + \frac{11\pi^2}{9} + 28\zeta_3 
 - \frac{16}{9\pi}\,
\int_0^1 \!dt_l \;\,
\frac{67-3\pi^2+33\ln(16t_l\bar t_l)}{\altsqrt{4t_l\bar t_l}}\;
\ln f(0,t_l) 
\nonumber\\[0.2em]
&\quad
+\frac{1}{\pi^2}\int_0^1 \!da \int_0^1 \!db \int_0^1 \!dt_l \int_0^1 \!dt_{kl} \;
\frac{k_{2}(a,b,t_{kl})}{\altsqrt{16t_l\bar t_l t_{kl}\bar t_{kl}}}\; 
\mathcal{F}(a,b,t_l,t_{kl})\,,
\label{eq:gamma1:nfca}
\end{align}
where $\bar t_{l}$ and $\bar t_{kl}$ are defined in analogy to $\bar t_{k}$, and
\begin{align}
\mathcal{F}(a,b,t_l,t_{kl}) &=
\ln \frac{F_A(a,b,0,t_k^+,t_l,t_{kl})}{f(0,t_l)} 
+\ln \frac{F_B(a,b,0,t_k^+,t_l,t_{kl})}{f(0,t_l)} 
+ (t_k^+\to t_k^-)
\end{align}
encodes the dependence on the two-emission measurement function. Here the subscripts $A$ 
and $B$ refer to two different versions of the measurement function with
\begin{align}
F_A(a,b,y,t_k,t_l,t_{kl}) &= F(a,b,y,t_k,t_l,t_{kl})\,,
\nonumber\\
F_B(a,b,y,t_k,t_l,t_{kl}) &= F(1/a,b,y,t_k,t_l,t_{kl})\,,
\end{align}
which arise because of certain remappings that are needed to constrain the integration 
region onto the unit hypercube~\cite{Bell:2018oqa}. We further introduced the angular variables
\begin{align}
t_k^\pm = t_l + t_{kl} - 2 t_l t_{kl} \pm 2 \sqrt{t_l\bar t_l t_{kl}\bar t_{kl}}
\end{align}
as well as the integration kernels 
\begin{align}
k_{1}(a,b,t_{kl}) &=
\frac{128a}{(a+b)^2(1+a b)^2} \;\bigg\{
\frac{b (1-a^2)^2}{[(1-a)^2+4a t_{kl}]^2}
- \frac{(a+b)(1+ab)}{(1-a)^2+4a t_{kl}} \bigg\}\,,
\nonumber\\[0.2em]
k_{2}(a,b,t_{kl}) &=
-\frac{32}{ab(a+b)^2(1+a b)^2} \;\bigg\{
\frac{2a^2b^2(1-a^2)^2}{[(1-a)^2+4a t_{kl}]^2}
- (a+b)(1+ab)
\nonumber\\[0.2em]
&\qquad\times
\bigg[ b(1+a^2)+2a(1+b^2) -\frac{b(1-a^2)^2+2a(1+a^2)(1+b^2)}{(1-a)^2+4a t_{kl}}
\bigg] \bigg\}\,.
\end{align}

The third colour structure in (\ref{eq:gamma1:colour}) is only non-zero for observables that 
violate the non-Abelian exponentiation (NAE) theorem~\cite{Gatheral:1983cz,Frenkel:1984pz}. 
For observables that obey NAE, the two-emission measurement function factorises in Laplace 
space into a product of single-emission functions, and one easily verifies that 
$\gamma^{C_F}_1$ vanishes in this case. With our general ansatz (\ref{eq:measure:NNLO:unc}) 
in terms of a non-factorisable function $G(y_k,y_l,b,t_k,t_l,t_{kl})$, it is however 
non-trivial to show that $\gamma^{C_F}_1$ is zero for observables that obey NAE. Moreover, 
we find that the $1/\eps^2$ poles only cancel as predicted by the RGE \eqref{eq:RGE:SCET-1} 
if the constraint \eqref{eq:CF2:constaint} in Appendix~\ref{app:CF2:details} is satisfied. 
For further details on the calculation of the $C_F^2$ contribution we refer to the appendix, 
but we stress once more that the following result for $\gamma^{C_F}_1$ only holds if the 
condition \eqref{eq:CF2:constaint} is fulfilled. Explicitly, we find
\begin{align}
\gamma^{C_F}_1 &=
 \frac{128}{\pi}\,
\int_0^1 \!dy \int_0^1 \!dt_l \;\,
\frac{1}{\altsqrt{4t_l\bar t_l}}\;\,\frac{1}{y}\;
\ln^2 \left( \frac{(1+y)^n\, f(y,t_l)}{f(0,t_l)} \right) 
\nonumber\\[0.2em]
&\quad
+\frac{256}{\pi}\,
\int_0^1 \!dy \int_0^1 \!dt_l \;\,
\frac{\ln f(0,t_l)}{\altsqrt{4t_l\bar t_l}}\;
 \frac{\ln f(y,t_l)}{y_+} 
\nonumber\\[0.2em]
&\quad
-\frac{512}{\pi^2}\,
\int_0^1 \!dt_k \;\,
\frac{\ln f(0,t_k)}{\altsqrt{4t_k\bar t_k}}\;
\int_0^1 \!dy \int_0^1 \!dt_l \;\,
\frac{1}{\altsqrt{4t_l\bar t_l}} \; 
\frac{\ln f(y,t_l)}{y_+}
\nonumber\\[0.2em]
&\quad
-\frac{128}{\pi^2}
\int_0^1 \!dy \int_0^1 \!db \int_0^1 \!dt_l \int_0^1 \!dt_{kl} \;
\frac{1}{\altsqrt{16t_l\bar t_l t_{kl}\bar t_{kl}}}\; \,
\frac{\mathcal{G}_1(y,b,t_l,t_{kl})}{y_+b_+}\,
\nonumber\\[0.3em]
&\quad
-\frac{64}{\pi^2}
\int_0^1 \!dr \int_0^1 \!db \int_0^1 \!dt_l \int_0^1 \!dt_{kl} \;
\frac{1}{\altsqrt{16t_l\bar t_l t_{kl}\bar t_{kl}}}\; \,
\frac{\mathcal{G}_2(r,b,t_l,t_{kl})}{r_+b_+}
\label{eq:gamma1:cf2}
\end{align}
up to an additional contribution in \eqref{eq:CF2:deltagamma1}, which we conjecture to 
vanish for all observables. Here the notation $1/x_+$ refers to a plus-distribution
defined as $\int_0^1 \!dx \,f(x)/x_+ =\int_0^1 \!dx \,(f(x)-f(0))/x$. 
The dependence on the two-emission measurement function is furthermore now encoded in
\begin{align}
\mathcal{G}_{1}(y,b,t_l,t_{kl}) &=
\ln G_{A_1}(y,0,b,t_k^+,t_l,t_{kl})
+ \ln G_{A_2}(y,0,b,t_k^+,t_l,t_{kl})
\nonumber\\
&\quad
+ \ln G_{B_1}(y,0,b,t_k^+,t_l,t_{kl})
+ \ln G_{B_2}(y,0,b,t_k^+,t_l,t_{kl})
+ (t_k^+\to t_k^-)\,,
\nonumber\\
\mathcal{G}_{2}(r,b,t_l,t_{kl}) &=
\ln G_{A_1}(0,r,b,t_k^+,t_l,t_{kl})
+ \ln G_{A_2}(0,r,b,t_k^+,t_l,t_{kl})
\nonumber\\
&\quad
+ \ln G_{B_1}(0,r,b,t_k^+,t_l,t_{kl})
+ \ln G_{B_2}(0,r,b,t_k^+,t_l,t_{kl})
+ (t_k^+\to t_k^-)\,,
\label{eq:G1G2}
\end{align} 
where the subscripts $A$ and $B$ refer to the same and opposite hemisphere contributions, 
respectively, with
\begin{align}
G_A(y_k,y_l,b,t_k,t_l,t_{kl}) &= G(y_k,y_l,b,t_k,t_l,t_{kl})\,,
\nonumber\\
G_B(y_k,y_l,b,t_k,t_l,t_{kl}) &= y_l^{-n} \;G(y_k,1/y_l,b,t_k,t_l,t_{kl})\,,
\end{align}
and we have disentangled the scalings in the joint limit $y_k\to0$ and $y_l\to0$ at a 
fixed ratio $y_k/y_l$ from those of the subsequent limits with $y_k/y_l\to0$ or 
$y_l/y_k\to0$ via
\begin{align}
G_{A_1}(y,r,b,t_k,t_l,t_{kl}) &= G_A(y,r y,b,t_k,t_l,t_{kl})\,,
\nonumber\\
G_{A_2}(y,r,b,t_k,t_l,t_{kl}) &= G_A(r y,y,b,t_k,t_l,t_{kl})\,,
\end{align}
and similarly for region $B$.

\begin{figure}
\begin{center}
\includegraphics[height=5.8cm]{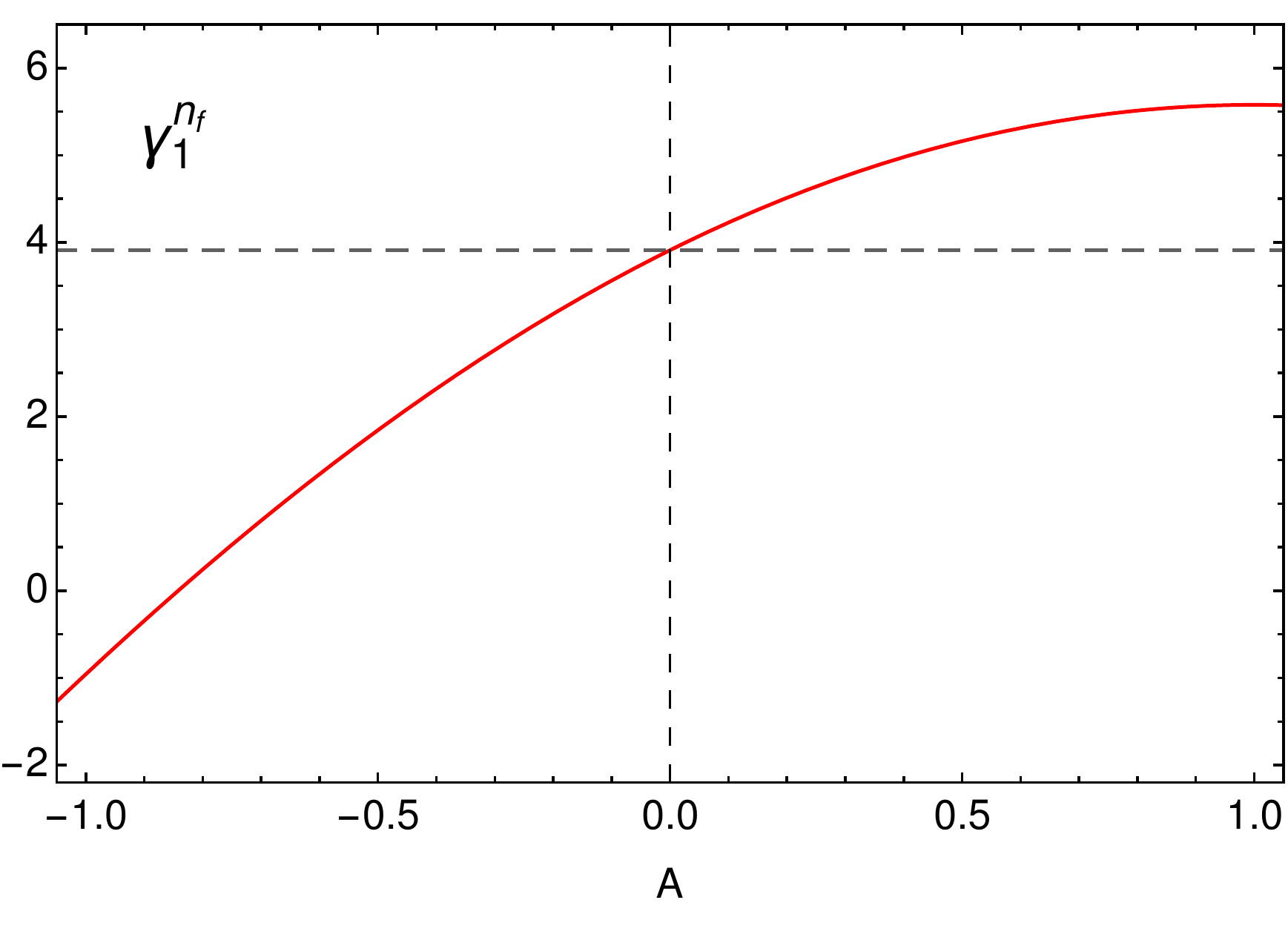}
\hspace{5mm}
\includegraphics[height=5.8cm]{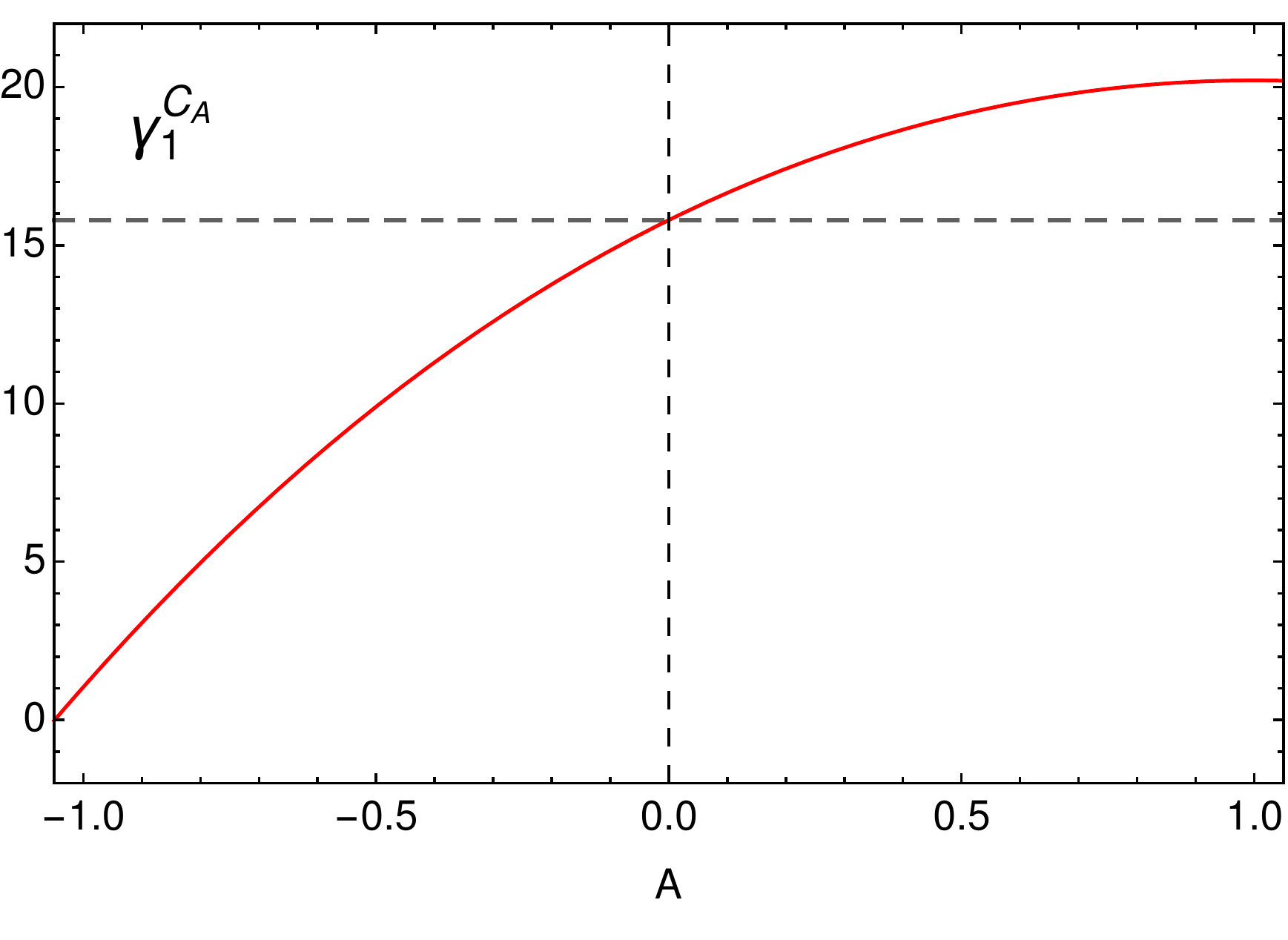}
\end{center}
\vspace{-0.7cm}
\caption{\label{fig:angularities}
Two-loop soft anomalous dimension of the $e^+e^-$ event-shape variable angularities. 
The dashed line indicates the thrust number, which is known analytically 
from~\cite{Kelley:2011ng,Monni:2011gb}.}
\end{figure}

Equations \eqref{eq:gamma0}, \eqref{eq:gamma1:nfca} and \eqref{eq:gamma1:cf2} represent the
main result of this section; they directly yield the soft anomalous dimension once the 
measure\-ment functions for an observable have been determined. Let us now illustrate how 
to use these equations with a few examples. For simplicity, we focus here on soft functions 
that obey NAE such that $\gamma^{C_F}_1=0$ in the following. We will come back to 
the discussion of NAE-violating observables in Section~\ref{sec:generalisation}.

We first consider the soft function relevant for threshold resummation in Drell-Yan 
production~\cite{Belitsky:1998tc,Becher:2007ty}, which is characterised by $n=-1$, 
$f(y_k,t_k) =1+y_k$ and $F(a,b,y,t_k,t_l,t_{kl}) =1+y$. It turns out that the integrals 
in \eqref{eq:gamma0} and \eqref{eq:gamma1:nfca} vanish for this observable and so 
$\gamma^{S}_0 =0$, $\gamma^{n_f}_1 =224/27 - 4/9\,\pi^2$ and
$\gamma^{C_A}_1 =-808/27 + 11/9\,\pi^2 + 28\zeta_3$, which agrees with the findings 
from~\cite{Belitsky:1998tc,Becher:2007ty}. With the explicit formulae for the measurement 
function from the previous section, one similarly shows that the soft anomalous 
dimension for $W$-production at large transverse momentum is identical (in our normalisation), 
which is in line with the calculation in~\cite{Becher:2012za}. The same is true for certain 
event-shape variables like thrust and 
C-parameter~\cite{Kelley:2011ng,Monni:2011gb,Hoang:2014wka}.

As a new application of our formalism, we consider the $e^+ e^-$ event shape 
angularities~\cite{Berger:2003iw}. In this case one has 
$n=1-A$, $f(y_k,t_k) =1$ (for $0\leq y_k\leq1$), and
\begin{align}
F_A(a,b,0,t_k,t_l,t_{kl}) &=\frac{a + a^A b}{a + b} \;
\bigg(\frac{a + b}{a(1 + a b)} \bigg)^{A/2}\,,
\nonumber\\
F_B(a,b,0,t_k,t_l,t_{kl}) &=\frac{a^A + a b}{1 + a b} \;
\bigg(\frac{1 +a b}{a(a + b)} \bigg)^{A/2}\,,
\end{align}
where $A<1$ is the value of the angularity\footnote{The precise definition of the 
angularities is given in \eqref{def:angularities} below. The value $A=0$ then corresponds to thrust
and $A=1$ is the total jet broadening (without recoil effects).}. At NLO this again implies 
$\gamma^{S}_0 =0$ in agreement with~\cite{Hornig:2009vb} and at NNLO the integral 
representations in \eqref{eq:gamma1:nfca} can be evaluated numerically. The result is shown 
in Figure~\ref{fig:angularities}, which represents the first calculation of the two-loop 
soft anomalous dimension for this observable. Our result can be used to extend existing 
resummations for the angularity distributions to NNLL 
accuracy~\cite{Bell:2017wvi,Procura:2018zpn,Bell:2018gce}.

\section{Collinear anomaly exponent}
\label{sec:SCET-2}

For observables with $n=0$, the phase-space integrals are sensitive to rapidity divergences
and the soft function is defined in SCET-2. In this case we implement the phase-space
regulator $\alpha$ as discussed at the end of Section~\ref{sec:measurement}, and we determine 
the collinear anomaly exponent $\mathcal{F}(\tau,\mu)$ from the $1/\alpha$ poles of the bare 
soft function. The collinear anomaly exponent controls the logarithmic corrections in the 
rapidity scale $\nu$~\cite{Becher:2010tm},
\begin{align}
S(\tau,\mu,\nu)
&= (\nu^2\taubar^2)^{-\mathcal{F}(\tau,\mu)} \;W_S(\tau,\mu)\,,
\end{align}
which can also be viewed as the solution of a rapidity RGE~\cite{Chiu:2012ir}. The 
renormalised anomaly exponent satisfies the RGE
\begin{align}
\label{eq:RGE:SCET-2}
\frac{\rd}{\rd \ln\mu}  \; \mathcal{F}(\tau,\mu)
&= 2 \,\Gamma_{\mathrm{cusp}}(\alpha_s)\,,
\end{align}
which has the two-loop solution
\begin{align}
\mathcal{F}(\tau,\mu) &= 
\left( \frac{\alpha_s}{4 \pi} \right) 
\Big\{ 2\Gamma_0 \,L 
+ d_1 \Big\}
+\left( \frac{\alpha_s}{4 \pi} \right)^2 
\Big\{ 2 \beta_0\Gamma_0\, L^2 
 + 2 \left( \Gamma_1 + \beta_0 d_1 \right) L + d_2 \Big\},
 \label{eq:d1d2}
\end{align}
where $L=\ln(\mu\taubar)$ and $\beta_0=11/3\,C_A-4/3\,T_Fn_f$ is the one-loop coefficient
of the beta function.

We then proceed along the lines of the previous section to determine the non-logarithmic
terms of the collinear anomaly exponent $d_1$ and $d_2$. At NLO we find that the one-loop
anomaly exponent has a similar integral representation as in \eqref{eq:gamma0} with
\begin{align} 
d_1 = -\gamma^{S}_0\,.
\label{eq:d1}
\end{align}
At NNLO we again distinguish between correlated and uncorrelated emission contributions,
and we decompose the two-loop anomaly exponent according to three colour structures,
\begin{align} 
d_2 &= d_2^{\,n_f} \,C_F T_F n_f
\,+\, d_2^{\,C_A} \,C_F C_A 
\,+\,  d_2^{\,C_F} \,C_F^2\,.
\end{align}
Intriguingly, we again find similar integral representations as in \eqref{eq:gamma1:nfca} 
and \eqref{eq:gamma1:cf2} with
\begin{align}
d_2^{\,n_f} &= -\gamma^{n_f}_1 - \frac{4\pi^2}{3} 
 - \frac{64}{3\pi}\,
\int_0^1 \!dt_l \;\,
\frac{\ln f(0,t_l) }{\altsqrt{4t_l\bar t_l}}\;
\ln \bigg(\frac{f(0,t_l)}{16t_l\bar t_l}\bigg)\,,
\nonumber\\
d_2^{\,C_A} &= -\gamma^{C_A}_1 + \frac{11\pi^2}{3} 
 + \frac{176}{3\pi}\,
\int_0^1 \!dt_l \;\,
\frac{\ln f(0,t_l) }{\altsqrt{4t_l\bar t_l}}\;
\ln \bigg(\frac{f(0,t_l)}{16t_l\bar t_l}\bigg)\,,
\nonumber\\
d_2^{\,C_F} &= -\gamma^{C_F}_1\,,
\label{eq:d2}
\end{align}
where the same remarks apply for the $C_F^2$ contribution as in the previous section,
namely our result for $d_2^{\,C_F}$ only holds if the constraint \eqref{eq:CF2:constaint} is 
satisfied and there exists an additional contribution to $d_2^{\,C_F}$ that is specified in 
\eqref{eq:CF2:deltad2} and which we conjecture to vanish for all observables.

We again illustrate the use of equations \eqref{eq:d1} and \eqref{eq:d2} with a few examples 
that respect NAE such that $d_2^{\,C_F} =0$. We first consider the soft function for the 
jet broadening event shape, neglecting any complications from recoil 
effects\footnote{A recoil-free definition of jet broadening was introduced 
in~\cite{Larkoski:2014uqa}, but we prefer to discuss the standard thrust-axis definition here 
since this allows us to compare our results with the NNLO calculation in~\cite{Becher:2012qc}.
By setting the variable $z$ to zero in this paper, the recoil effects can be switched off 
and the soft function can then be written in the form \eqref{eq:recoilfreebroad}.}. It is 
specified by $f(y_k,t_k) =1/2$ and
\begin{align}\label{eq:recoilfreebroad}
F(a,b,y,t_k,t_l,t_{kl}) &=
\sqrt{\frac{a}{(1+a b)(a+b)}}\;
\frac{1+b}{2}\,,
\end{align}
which yields $d_1 =-8C_F\ln2$. The two-loop anomaly exponent can also be obtained analytically 
in our setup, and we find
\begin{align}
d_2^{\,n_f} &=-\frac{32}{3} \ln^2 2 + \frac{320}{9} \ln 2 - \frac{128}{27} 
- \frac{8 \pi^2}{3}\,,
\nonumber\\
d_2^{\,C_A} &=\frac{88}{3} \ln^2 2 - \left(\frac{1048}{9} + \frac{16 \pi^2}{3}\right) \ln 2 
+ \frac{760}{27} + \frac{22\pi^2}{3} + 8\zeta_3\,,
\end{align}
which agrees with the results from~\cite{Becher:2012qc}.

Another interesting application is the soft function for transverse-momentum resummation in
Drell-Yan production~\cite{Echevarria:2015byo,Luebbert:2016itl}. In this case, one has 
$f(y_k,t_k) =-2 i (1 - 2 t_k)$ and
\begin{align}
F(a,b,y,t_k,t_l,t_{kl}) &=
-2 i \;\sqrt{\frac{a}{(1+a b)(a+b)}}\; \Big(b (1 - 2 t_k) + 1 - 2 t_l\Big)\,,
\end{align}
where the global factor of $i$ arises from taking a Fourier instead of a Laplace 
transformation. Although the imaginary unit may lead to non-trivial phases for the 
individual terms in \eqref{eq:d2}, the imaginary parts must cancel in their sum since 
the anomaly exponent is real. The results from Sections~\ref{sec:SCET-1} and \ref{sec:SCET-2} 
can therefore equally be applied to soft functions that are defined in Fourier space, 
and for the specific case of transverse-momentum resummation, we find $d_1 =0$ at NLO,
and $d_2^{\,n_f}=-8.294(8)$ and $d_2^{\,C_A}=-3.727(11)$ at NNLO, 
which is in excellent agreement with the analytic results $d_2^{\,n_f} = -224/27$ 
and $d_2^{\,C_A} = 808/27-28\zeta_3$ from~\cite{Becher:2010tm,Gehrmann:2014yya}.

\section{Relation between $\gamma^S$ and $\mathcal{F}$}
\label{sec:SCET-1vsSCET-2}

The results of the previous section suggest that there exists a relation between the soft 
anomalous dimension $\gamma^S(\alpha_s)$ and the collinear anomaly exponent 
$\mathcal{F}(\tau,\mu)$, which we explore more generally in this section. As we are 
mainly interested in understanding the mismatch between $\gamma_1$ and $d_2$ that arises 
at NNLO in \eqref{eq:d2}, we focus in this section on observables that are consistent 
with NAE. For concreteness, we consider the angularity event shape~\cite{Berger:2003iw}
\begin{align}
\label{def:angularities}
  e_A(X) = \sum_{i\in X} \,|k_\perp^i| \; e^{-|\eta_i|\,(1-A)}\,,
\end{align}
where the transverse momentum $k_\perp^i$ and the rapidity $\eta_i$ are
measured with respect to the thrust axis. The angularities obey
a SCET-1 type factorisation theorem for $A<1$ in the dijet limit 
$e_A\ll1$~\cite{Hornig:2009vb}. The case $A=1$, on the other hand, corresponds to the 
event shape total jet broadening, which is a SCET-2 observable\footnote{We again neglect 
recoil effects in this section and one in addition has to account for a different 
normalisation of $e_1$ compared to the standard definition of jet 
broadening.‌}~\cite{Becher:2011pf,Chiu:2012ir}. In the limit $A\to 1$, we can thus examine 
the transition from SCET-1 to SCET-2 and in this way we can connect the soft anomalous 
dimension with the collinear anomaly exponent. Our analysis is inspired by and extends 
the study of~\cite{Larkoski:2014uqa}. 

The starting point of our analysis is the resummed angularity distribution in Laplace
space,
\begin{align}
\frac{1}{\sigma_0} \frac{\mathrm{d}\sigma}{\mathrm{d}\tau_A}&=
e^{4\mathcal{S}(\mu_h,\mu_j) -2A_H(\mu_h,\mu_j) 
+\frac{4}{1-A}\mathcal{S}(\mu_s,\mu_j)+\frac{2}{1-A}A_S(\mu_j,\mu_s)}\;
 \left(\frac{Q^2}{\mu_h^2}\right)^{-2A_\Gamma(\mu_h,\mu_j)}\;
\left(\mu_s \bar \tau_A\right)^{-\frac{4}{1-A}A_\Gamma(\mu_j,\mu_s)}
\nonumber\\
&\quad\times
H(Q,\mu_h) \;J( \tau_A,\mu_j)\; J( \tau_A,\mu_j) \; S( \tau_A,\mu_s)\,,
\label{eq:scet1:resum}
\end{align}
where $\tau_A$ represents the Laplace-conjugate variable to $e_A$ and the scales $\mu_i$ 
with $i=h,j,s$ are to be chosen such that the quantities in the second line do not contain 
large logarithmic corrections. The evolution kernels
\begin{align}
  \mathcal{S}(\mu_1, \mu_2) &= - \int_{\alpha_s (\mu_1)}^{\alpha_s (\mu_2)} d\alpha\;
  \frac{\Gamma_{\mathrm{cusp}} (\alpha)}{\beta (\alpha)} \;\int_{\alpha_s
  (\mu_1)}^{\alpha} \,\frac{d\alpha'}{\beta (\alpha')}\,,
	\nonumber\\
  A_i(\mu_1, \mu_2) &= - \int_{\alpha_s (\mu_1)}^{\alpha_s (\mu_2)} d\alpha\;
  \frac{\gamma^i (\alpha)}{\beta (\alpha)}
\end{align}
for $i=H,S$ and $A_\Gamma(\mu_1, \mu_2)$, which is defined as $A_i(\mu_1, \mu_2)$ but 
with $\gamma^i$ replaced by $\Gamma_{\mathrm{cusp}}$, then resum the logarithmic 
corrections to all orders in perturbation theory.

Whereas the above expression holds for $A<1$, one can apply the collinear anomaly 
technique~\cite{Becher:2010tm} or, equivalently, the rapidity RG~\cite{Chiu:2012ir} to 
resum the logarithmic corrections to the (recoil-free) broadening distribution in the 
dijet limit. In this case, one finds
\begin{align}
\frac{1}{\sigma_0} \frac{\mathrm{d}\sigma}{\mathrm{d}\tau_1}&=
e^{4\mathcal{S}(\mu_h,\mu_s) -2A_H(\mu_h,\mu_s)}\;
 \left(\frac{Q^2}{\mu_h^2}\right)^{-2A_\Gamma(\mu_h,\mu_j)}\;
 \left(\frac{\nu_j^2}{\nu_s^2}\right)^{-\mathcal{F}(\tau_1,\mu_s)}
\nonumber\\
&\quad\times
H(Q,\mu_h) \;J( \tau_1,\mu_j,\nu_j)\; J( \tau_1,\mu_j,\nu_j) \; 
S( \tau_1,\mu_s,\nu_s)\,,
\label{eq:scet2:resum}
\end{align}
where $\mathcal{F}(\tau_1,\mu_s)$ is the collinear anomaly exponent and $\nu_j$ and 
$\nu_s$ are rapidity scales. Notice that in our notation we distinguish between the 
SCET-2 jet and soft functions and the corresponding ones in \eqref{eq:scet1:resum} only 
by their arguments.

Our goal thus consists in connecting equations \eqref{eq:scet1:resum} and 
\eqref{eq:scet2:resum} in the limit $A\to1$. To this end, we first compare the terms that 
resum the double-logarithmic corrections in the RGEs of the respective soft functions,
\begin{align}
\frac{\rd\ln S(\tau_A,\mu_s)}{\rd \ln\mu_s}
&= - \frac{4}{1-A} \;\Gamma_{\mathrm{cusp}}(\alpha_s) \ln(\mu_s\taubar_A) 
+\ldots   \,,
\nonumber\\
\frac{\rd\ln S(\tau_1,\mu_s,\nu_s)}{\rd \ln\mu_s}
&= 4 \,\Gamma_{\mathrm{cusp}}(\alpha_s) \ln(\mu_s\taubar_1) 
-4 \,\Gamma_{\mathrm{cusp}}(\alpha_s) \ln(\nu_s\taubar_1)
+\ldots   \,.
\end{align}
As the two equations must coincide in the limit $A\to1$, we obtain a relation between the 
soft RG scale $\mu_s$ and the corresponding rapidity scale $\nu_s$,
\begin{align}
\mu_s &= \nu_s^{\frac{1-A}{2-A}}\; \bar{\tau}_1^{-\frac{1}{2-A}}\,.
\end{align}
Proceeding similarly for the jet functions, one obtains
\begin{align}
\frac{\mu_j}{\mu_s} &= 
\left(\frac{\nu_j}{\nu_s}\right)^{\frac{1-A}{2-A}}
\,\xlongrightarrow{A\to1\;}\, 
1 + (1-A) \,\ln \frac{\nu_j}{\nu_s} + \mathcal{O}(1-A)^2\,,
\end{align}
which reveals that the jet and soft RG scales coincide in the strict broadening limit, 
and that the rapidity logarithms emerge in the first-order correction.

The preceding relation can be used to analyse the RG kernels in \eqref{eq:scet1:resum} 
that are multiplied by a divergent prefactor in the broadening limit. This yields
\begin{align}
\frac{4}{1-A}\,\,\mathcal{S}(\mu_s,\mu_j) &\,\xlongrightarrow{A\to1\;}\,
\mathcal{O}(1-A)\,,
\nonumber\\
\frac{2}{1-A}\;A_S(\mu_j,\mu_s) &\,\xlongrightarrow{A\to1\;}\,
\gamma^{S}\big(\alpha_s(1/\bar\tau_1)\big) 
\ln\frac{\nu_j^2}{\nu_s^2} + \mathcal{O}(1-A)\,,
\end{align}
which shows that the rapidity logarithms are indeed controlled by the soft anomalous 
dimension. In order to connect equations \eqref{eq:scet1:resum} and \eqref{eq:scet2:resum}, 
we still need, however, to examine the matching corrections more closely.

It turns out that the SCET-1 matching corrections are by themselves divergent in the limit
$A\to1$. Up to the considered NNLL accuracy, they can be written in the form
\begin{align}
J( \tau_A,\mu_j) &\,\xlongrightarrow{A\to1\;}\,
1 + \frac{\alpha_s(\mu_j)}{4 \pi} \bigg\{
\gamma_0^S \,\ln \frac{Q}{\nu_j} + \frac{d'_1}{2(1-A)} + c_1^J + \mathcal{O}(1-A) \bigg\}\,,
\nonumber\\
S( \tau_A,\mu_s) &\,\xlongrightarrow{A\to1\;}\,
1 + \frac{\alpha_s(\mu_s)}{4 \pi} \bigg\{
2\gamma_0^S \,\ln (\nu_s \bar\tau_1) - \frac{d'_1}{(1-A)}+ c_1^S + \mathcal{O}(1-A) \bigg\}\,.
\label{eq:scet1:matching}
\end{align}
Whereas the regular terms in the limit $A\to1$ match onto the corresponding expressions in 
the SCET-2 matching corrections (after identifying $d_1 = -\gamma_0^S$ as shown below), 
the pole terms in $(1-A)$ induce a new contribution. Although the poles themselves cancel 
in the product of jet and soft functions, their interplay generates a non-trivial correction 
since the coupling constants in \eqref{eq:scet1:matching} are evaluated at different scales. 
In total, we find that the ratio of SCET-1 and SCET-2 matching corrections becomes
\begin{align}
\frac{J( \tau_A,\mu_j)\; J( \tau_A,\mu_j) \; S( \tau_A,\mu_s)}{J( \tau_1,\mu_j,\nu_j)\; 
J( \tau_1,\mu_j,\nu_j) \; S( \tau_1,\mu_s,\nu_s)}
&\,\xlongrightarrow{A\to1\;}\,
1 - \left(\frac{\alpha_s(1/\bar\tau_1)}{4 \pi}\right)^2 \beta_0 \,d'_1 \,
\ln \frac{\nu_j^2}{\nu_s^2}\,,
\end{align}
which must be interpreted as an additional contribution to the anomaly exponent.

We now have assembled all pieces to combine equations \eqref{eq:scet1:resum} and 
\eqref{eq:scet2:resum} in the limit $A\to1$ and to connect the soft anomalous dimension 
with the collinear anomaly exponent. Up to NNLO, we find
\begin{align} 
d_1 &= -\gamma^{S}_0\,,
\nonumber\\
d_2 &= -\gamma^{S}_1 + \beta_0 \,d'_1\,,
\label{eq:d1d2:new}
\end{align}
which explains the relation and the mismatch between $d_2$ and $\gamma^{S}_1$ that we found 
earlier in \eqref{eq:d2}. The coefficient $d'_1$, which we introduced in 
\eqref{eq:scet1:matching} as the coefficient of the $(1-A)$ pole in the one-loop matching 
corrections, can furthermore be identified with the $\mathcal{O}(\eps)$ piece of the 
one-loop anomaly exponent,
\begin{align}
\mathcal{F}(\tau_1,\mu_s=1/\bar\tau_1) &= 
\left( \frac{\alpha_s}{4 \pi} \right) 
\Big\{d_1 + d'_1 \,\eps\Big\}
+ \mathcal{O}(\alpha_s^2)\,.
\end{align}

Our results in \eqref{eq:d1d2:new} resemble similar relations between the soft anomalous 
dimension and the collinear anomaly exponent that were found earlier 
in~\cite{Li:2016ctv,Vladimirov:2016dll}. The physical contents of these relations and our 
results is, however, different. Whereas the authors in~\cite{Li:2016ctv,Vladimirov:2016dll}
found a relation between the anomaly exponent for transverse-momentum resummation and the 
anomalous dimension for threshold resummation using either bootstrapping~\cite{Li:2016ctv} 
or conformal-mapping~\cite{Vladimirov:2016dll} techniques, our results assume that the same 
measurement functions $f(y_k,t_k)$, $F(a,b,y,t_k,t_l,t_{kl})$ and $G(y_k,y_l,b,t_k,t_l,t_{kl})$ 
enter the SCET-1 formulae \eqref{eq:gamma0}, \eqref{eq:gamma1:nfca}, \eqref{eq:gamma1:cf2} 
and the corresponding SCET-2 relations \eqref{eq:d1} and \eqref{eq:d2}. We therefore cannot 
connect the quantities for transverse-momentum and threshold resummation in our formalism, 
but in contrast our result can be used, for instance, to determine the 
(recoil-free) jet broadening anomaly exponent directly from the angularity soft anomalous 
dimension for $A=1$.

\section{Generalisation to other observables}
\label{sec:generalisation}

Our findings so far are limited to dijet soft functions with a measurement function 
that can be written in the form \eqref{eq:measure:NLO}, \eqref{eq:measure:NNLO:corr} and 
\eqref{eq:measure:NNLO:unc}. In this section we consider three types of generalisations: 
(i) cumulant soft functions with measurement functions that involve theta functions instead
of exponentials, (ii) multi-differential soft functions that depend on more than one Laplace
variable and (iii) $N$-jet soft functions that are defined in terms of $N>2$ light-like 
directions. We will address each of these extensions in turn.
\\[-0.5em]

\noindent
\textbf{Cumulant soft functions:} 
Soft functions for jet-veto and jet-grooming observables typically involve measurement
functions that are formulated in terms of a theta function, which reflects the fact that 
the jet veto/groomer provides a cutoff for the phase-space integrations of the soft radiation. 
Our formalism can easily be generalised to this class of observables. To do so, we write 
the one-emission measurement function of a cumulant soft function $\widehat{S}(\omega,\mu)$ 
in the form
\begin{align}
\widehat{\mathcal{M}}_1(\omega; k) = 
\theta\big(\omega - k_{T}\, y_k^{n/2}\, f(y_k,t_k)\,\big)\,,
\end{align}
and similarly for the two-emission functions. By taking the Laplace transform with respect
to the cutoff variable $\omega$, one can then bring the measurement function into the form 
considered in Section~\ref{sec:measurement}. We can thus calculate the bare soft function 
in Laplace space using the strategy from the previous sections, and we finally have to 
invert the Laplace transformation which reshuffles some of the coefficients in the $\eps$ and
$\alpha$ expansions. Assuming that the RGEs for cumulant soft functions $\widehat{S}(\omega,\mu)$
take a similar form 
as \eqref{eq:RGE:SCET-1} and \eqref{eq:RGE:SCET-2}, we can derive master formulae for the 
calculation of the soft anomalous dimension $\widehat{\gamma}^{S}(\alpha_s)$ and the 
anomaly exponent $\widehat{\mathcal{F}}(\omega,\mu)$ for this class of observables. 
Specifically, we find that the results in Section~\ref{sec:SCET-1} can be directly carried 
over for SCET-1 type cumulant soft functions,
\begin{align} 
\widehat{\gamma}^{S}_0 &= \gamma^{S}_0\,,
\nonumber\\
\widehat{\gamma}^{S}_1 &= \gamma^{S}_1\,,
\end{align}
but that the constraints for the $C_F^2$ contribution in Appendix~\ref{app:CF2:details} 
are slightly modified in this case. For SCET-2 type cumulant soft functions, we find that
the two-loop anomaly exponent receives an additional contribution proportional to 
$\beta_0=11/3\,C_A-4/3\,T_Fn_f$ with
\begin{align} 
\widehat{d}_1 &= d_1\,,
\nonumber\\
\widehat{d}_2 &= d_2
-\frac{\pi^2}{3} \, \beta_0\,\Gamma_0\,,
\end{align}
and the $C_F^2$ constraints are again slightly modified for these observables as 
discussed in more detail in Appendix~\ref{app:CF2:details}.

\begin{figure}[t]
\begin{center}
\includegraphics[height=3.9cm]{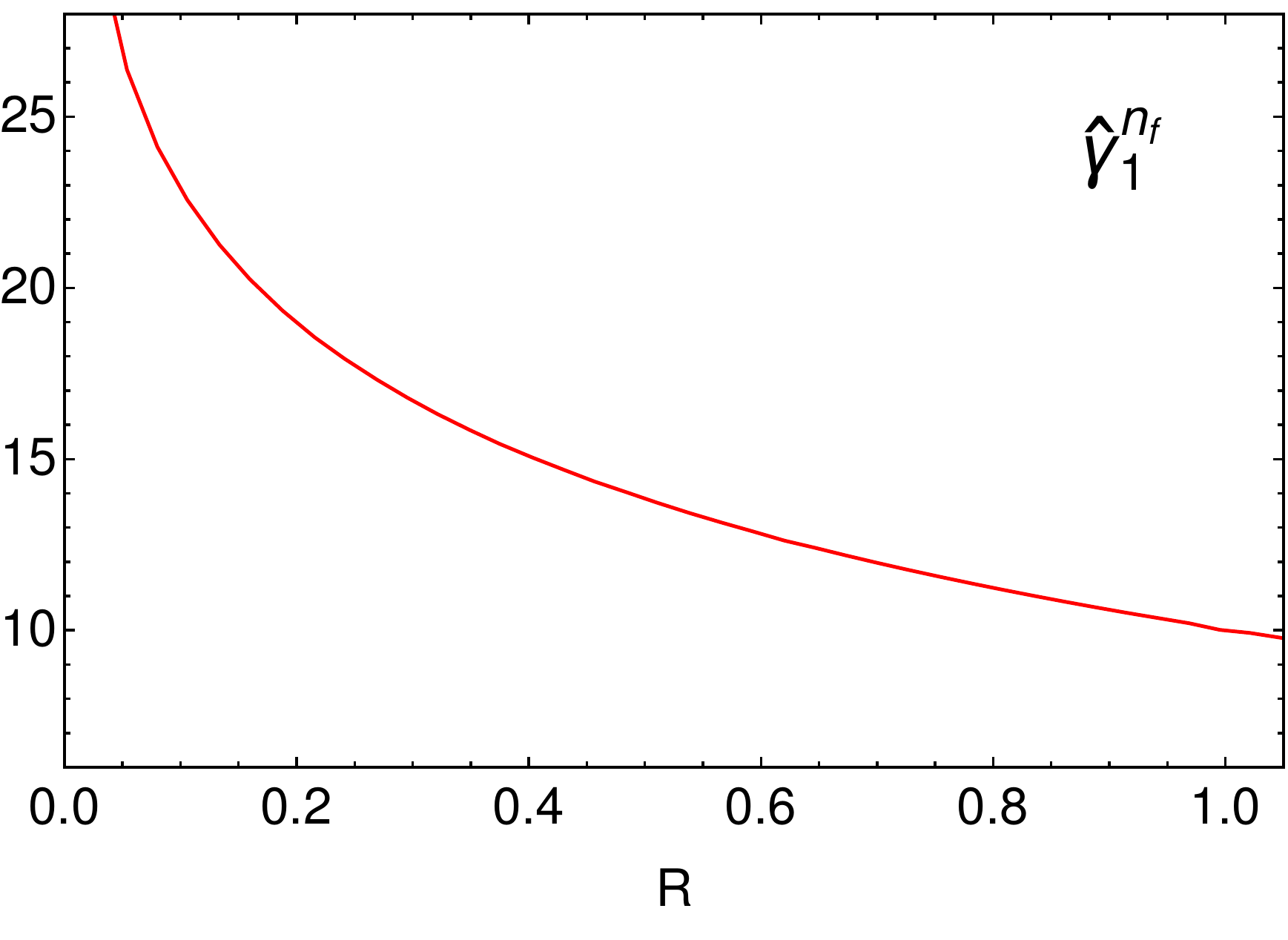}
\hspace{1.3mm}
\includegraphics[height=3.9cm]{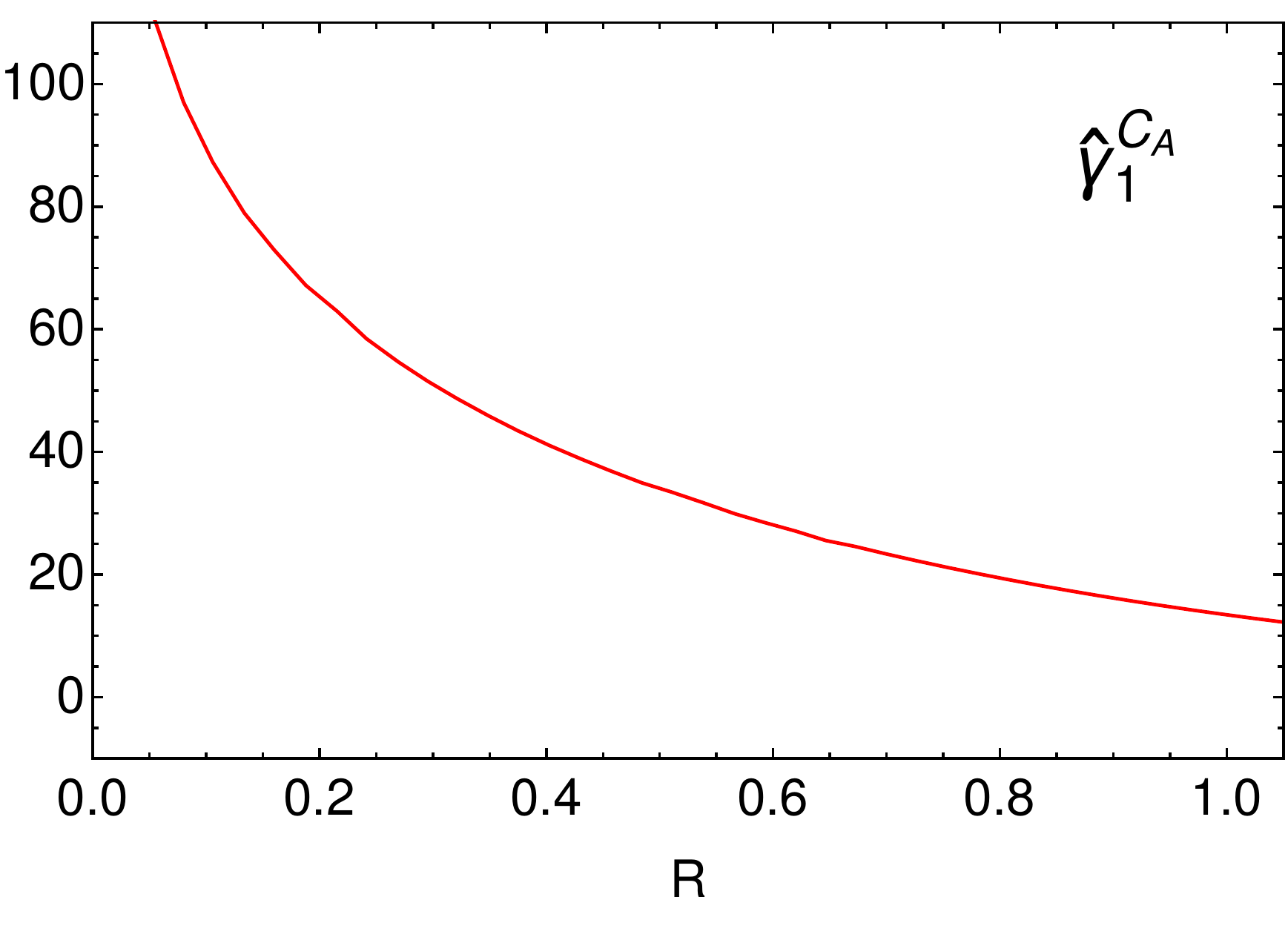}
\hspace{1.3mm}
\includegraphics[height=3.9cm]{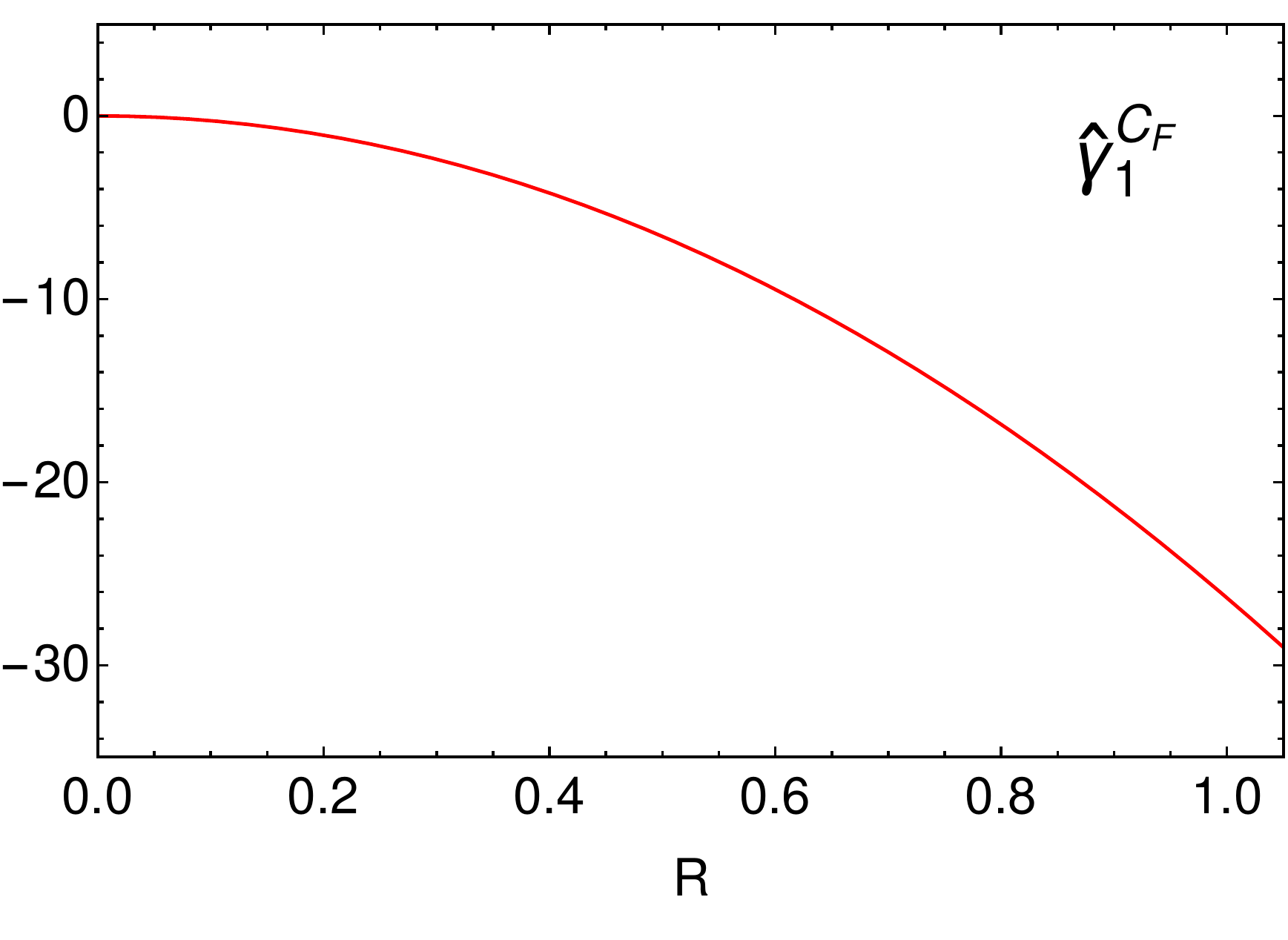}
\\[2mm]
\includegraphics[height=3.9cm]{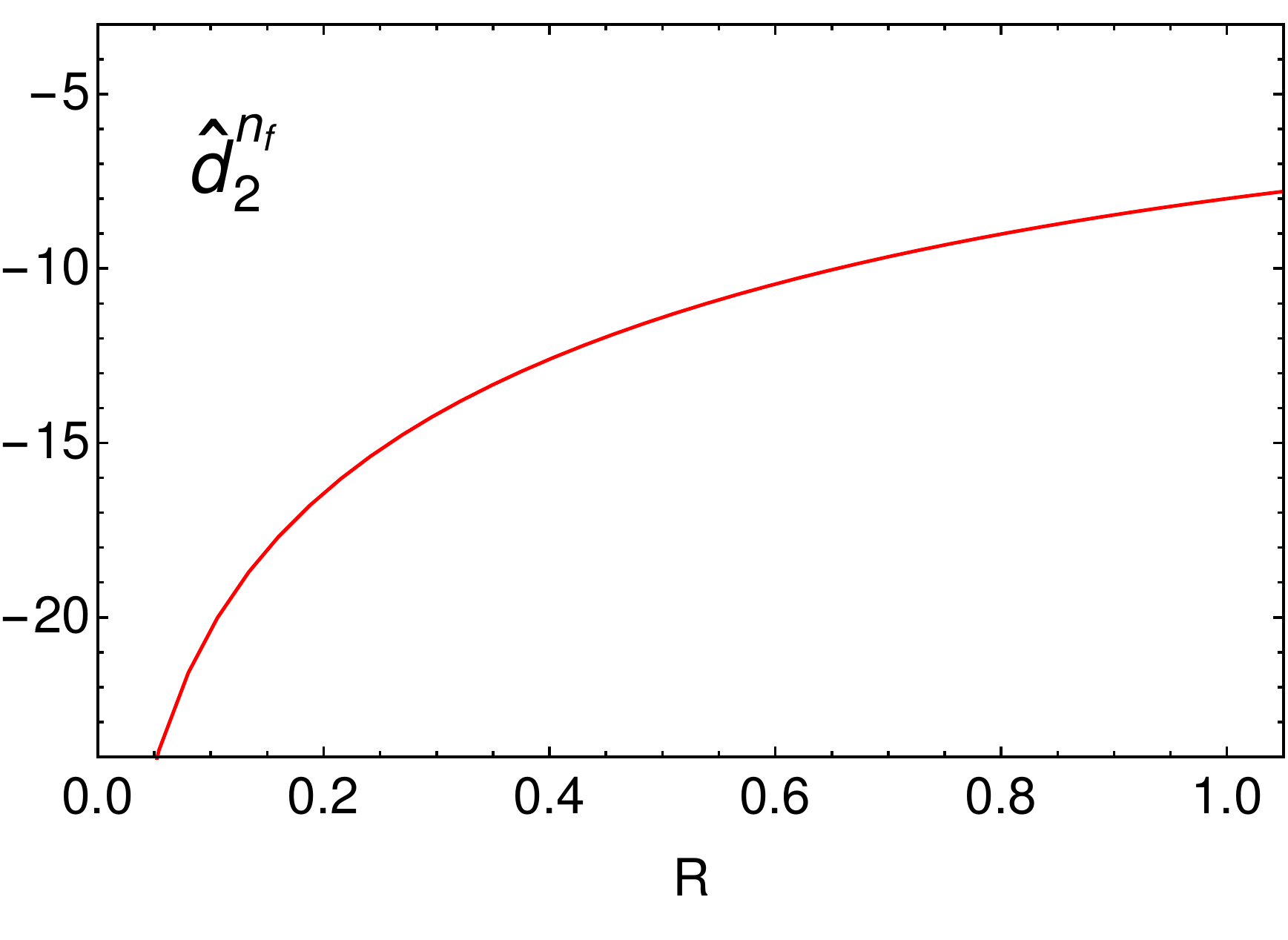}
\hspace{1.3mm}
\includegraphics[height=3.9cm]{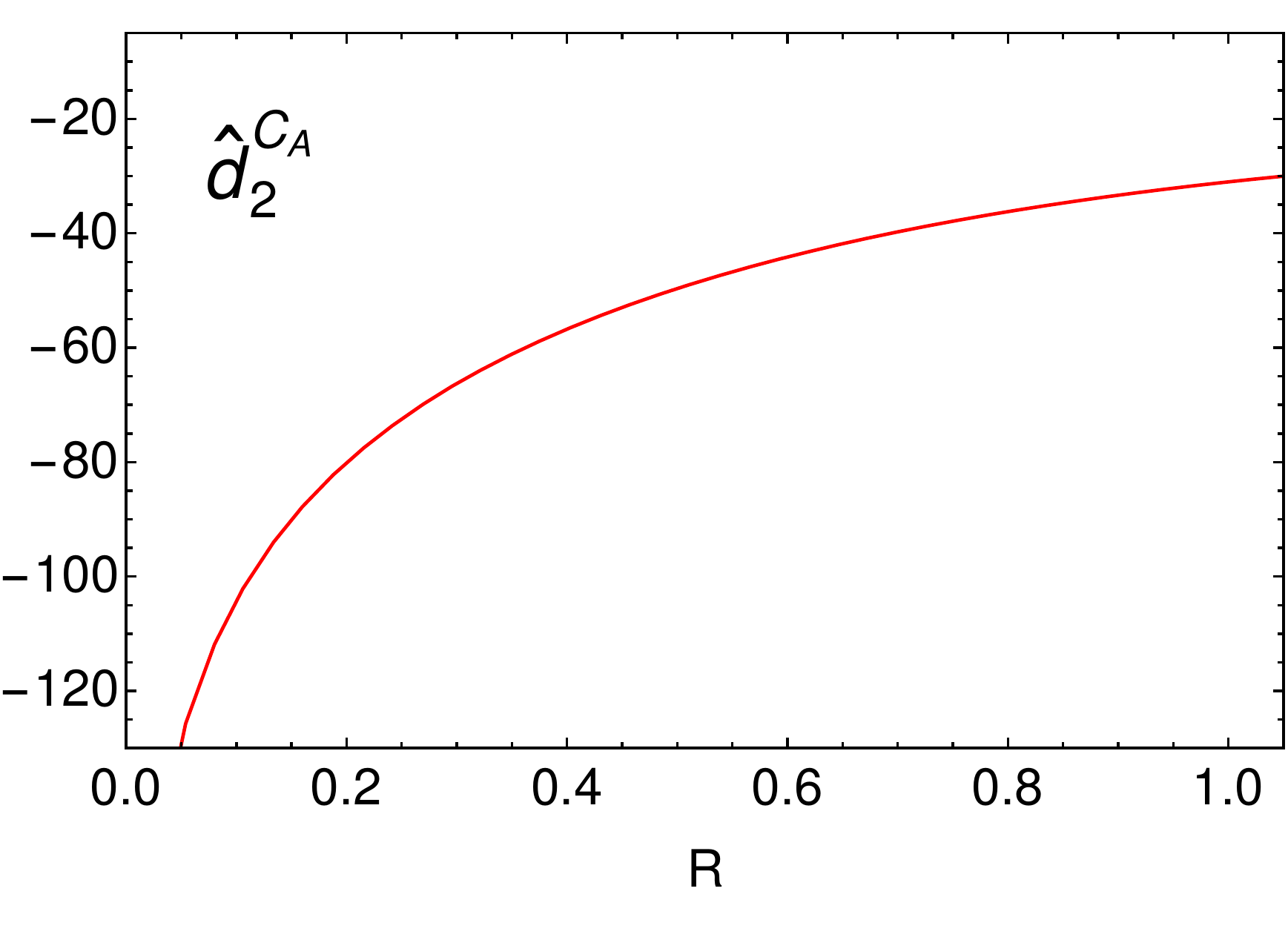}
\hspace{1.3mm}
\includegraphics[height=3.9cm]{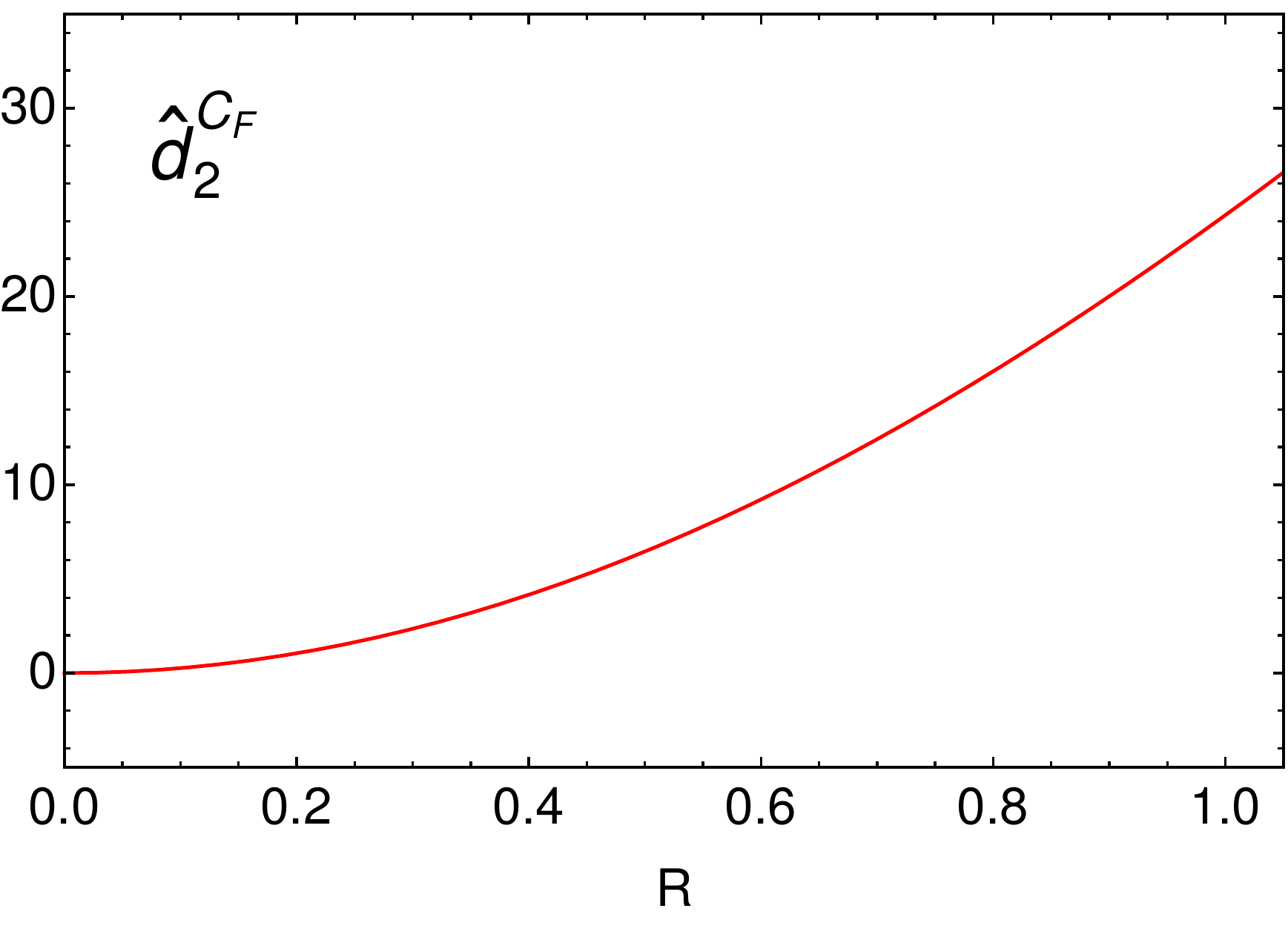}
\end{center}
\vspace{-0.5cm}
\caption{
Two-loop soft anomalous dimension of the rapidity-dependent jet-veto observables  
(upper row) and two-loop anomaly exponent of the $p_T$ 
jet veto (lower row).
\label{fig:jetvetoes}}
\end{figure}

In order to test the validity of these equations, we consider the soft functions for 
the rapidity-dependent jet-veto observables from~\cite{Gangal:2014qda} and the standard 
transverse-momentum based jet veto from~\cite{Becher:2013xia,Stewart:2013faa}. With the 
explicit form of the measurement functions from Appendix~\ref{app:cumulant:details}, we can 
then determine the soft anomalous dimension for the former and the collinear anomaly exponent 
for the latter. At NLO, this yields $\widehat{\gamma}^{S}_0=0$ and $\widehat{d}_1=0$, 
respectively, and at NNLO our results are shown as a function of the jet radius $R$ in 
Figure~\ref{fig:jetvetoes}. We compared these curves to the interpolating functions 
provided in~\cite{Gangal:2016kuo} for the rapidity-dependent jet vetoes and 
in~\cite{Banfi:2012yh,Becher:2013xia,Stewart:2013faa} for the $p_T$ veto and found agreement 
for all colour structures (the difference between these functions and our results is not 
visible on the scale of the plots). In particular, this represents the first validation 
of the expression in \eqref{eq:gamma1:cf2} and the last equation in \eqref{eq:d2}, which are 
needed only for observables that violate the NAE theorem.

\begin{figure}[t]
\begin{center}
\includegraphics[width=0.318\textwidth]{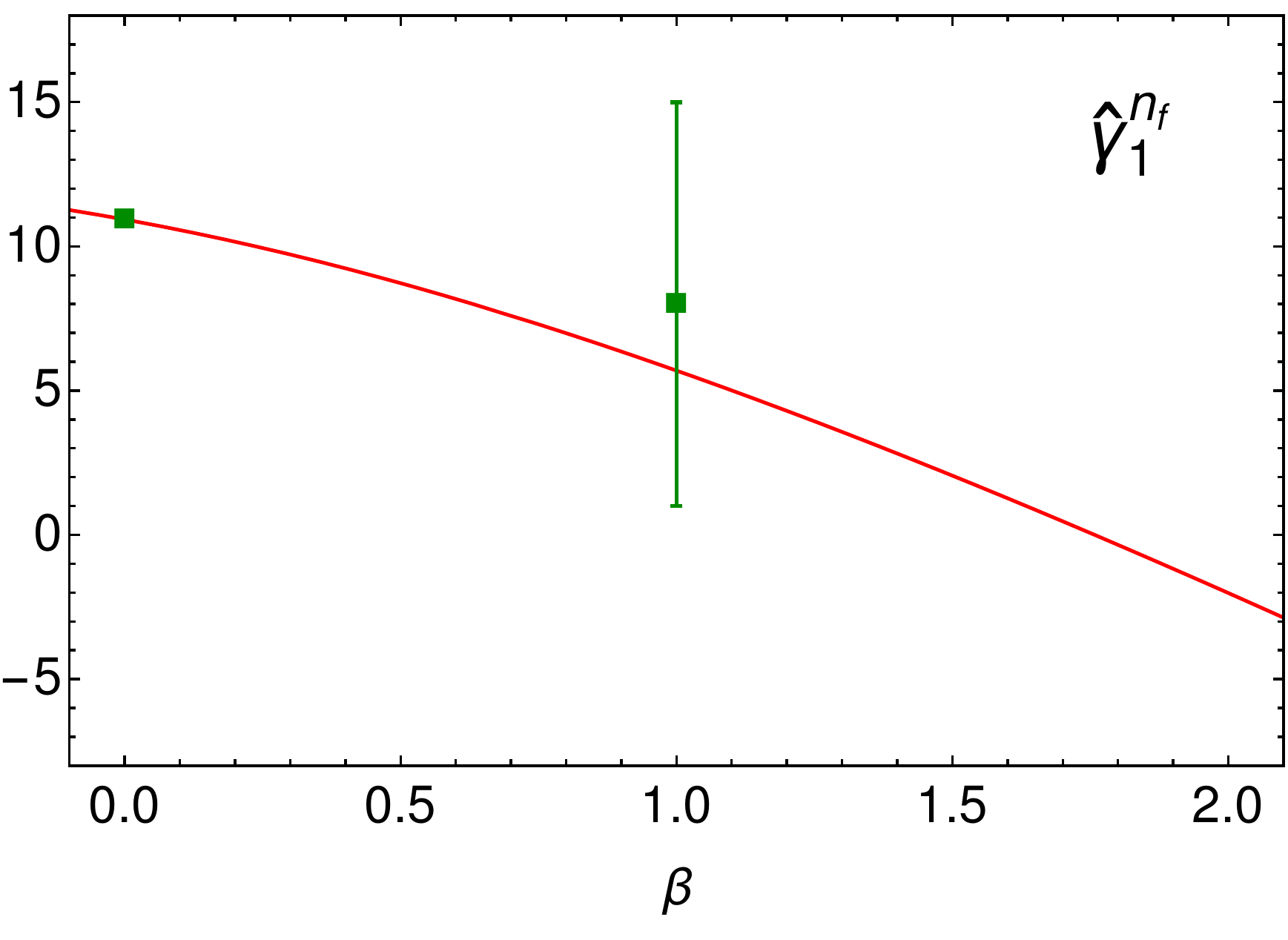}
\hspace{1.0mm}
\includegraphics[width=0.318\textwidth]{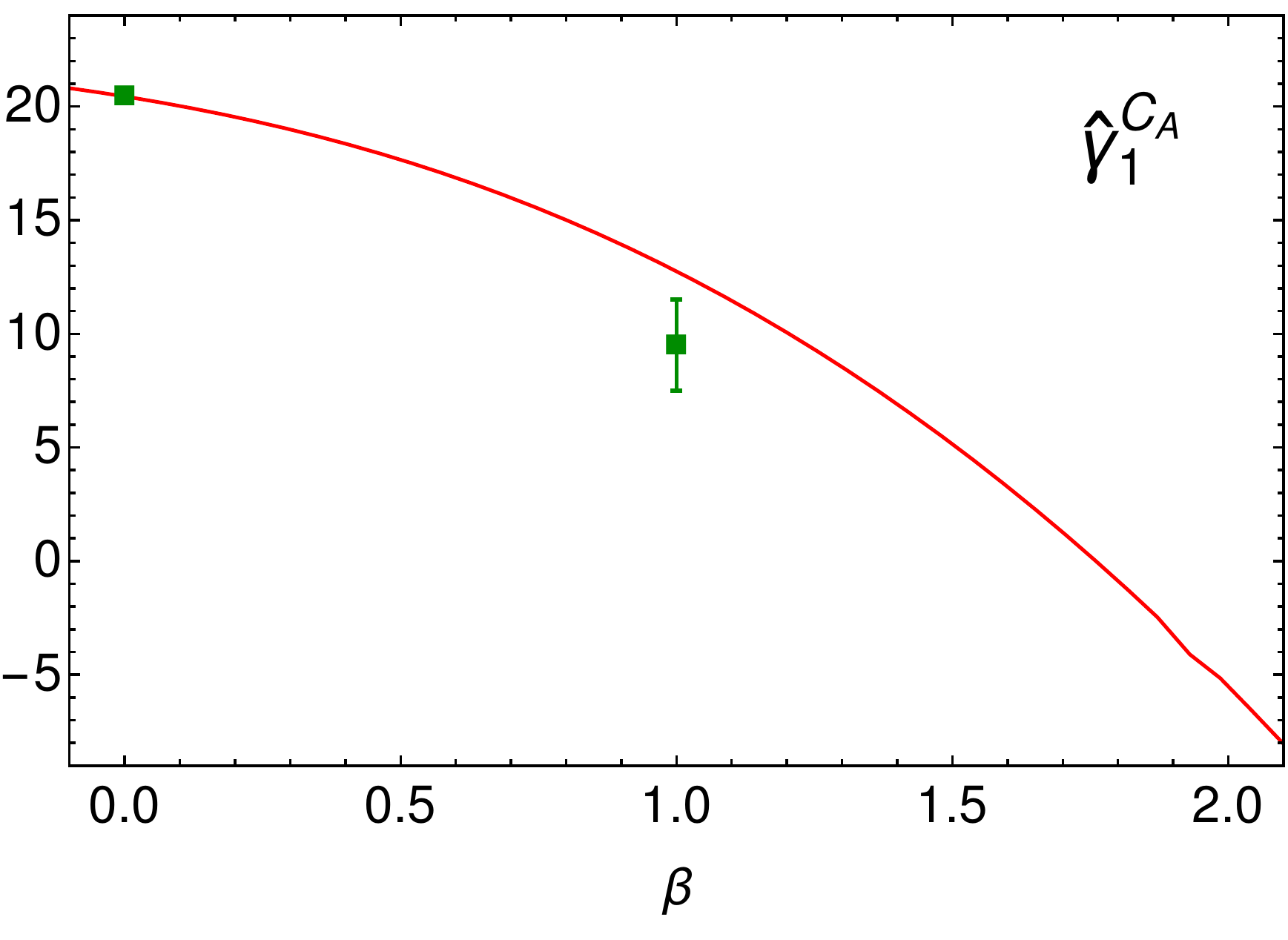}
\hspace{1.0mm}
\includegraphics[width=0.318\textwidth]{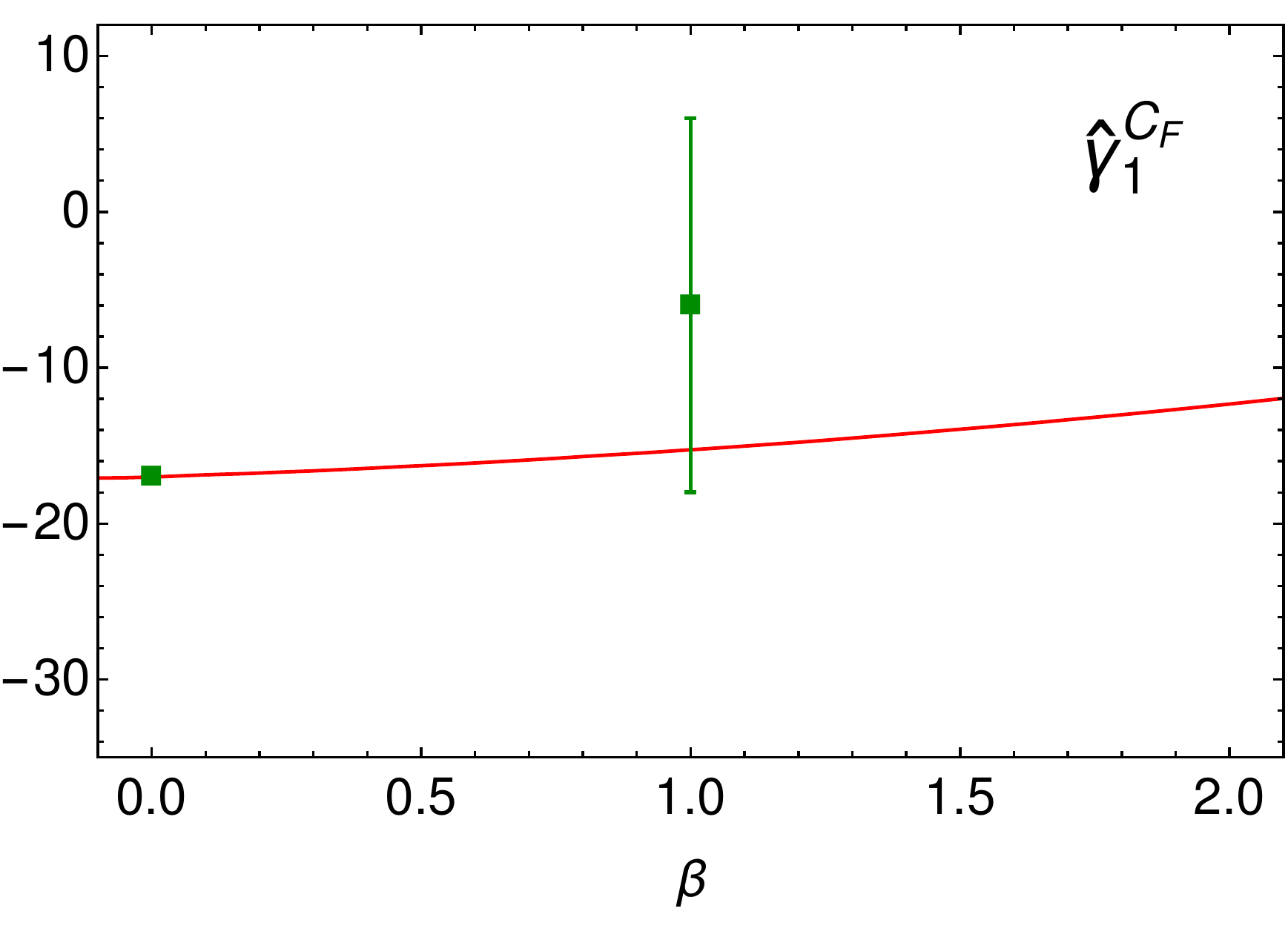}
\end{center}
\vspace{-0.5cm}
\caption{
Two-loop soft anomalous dimension of the soft-drop jet groomer. The (green) squares with 
error bars represent the results of~\cite{Frye:2016aiz} --- see the text.
\label{fig:jetgrooming}}
\end{figure}

As a new application of our formalism, we consider the soft function for the soft-drop 
jet-grooming algorithm from~\cite{Frye:2016aiz}. In this case we have $n= -1-\beta$, where 
$\beta$ is a parameter that controls the aggressiveness of the jet groomer (the explicit 
expressions for the measurement function can be found in Appendix~\ref{app:cumulant:details}). 
For values of $\beta>0$ considered here, the soft function is thus defined in SCET-1, and at 
NLO one finds $\widehat{\gamma}^{S}_0=0$. At NNLO our results for the soft anomalous 
dimension are displayed in Figure~\ref{fig:jetgrooming} as a function of the grooming
parameter $\beta$. The plots also show the numbers of an analytic extraction for $\beta=0$ 
and an {\tt EVENT2} fit for $\beta=1$ from~\cite{Frye:2016aiz}. As can be seen from the plots, 
we confirm the value for $\beta=0$, but our numbers for $\beta=1$ are by far more precise 
than the ones from~\cite{Frye:2016aiz}. The two-loop soft anomalous dimension has not been 
determined for other values of $\beta$ before.\\[-0.5em]

\noindent
\textbf{Multi-differential soft functions:}
Soft functions for exclusive observables typically depend on more than one kinematic 
variable such that several Laplace transformations may be needed to resolve all 
distributions. In this case, we choose the first Laplace variable $\tau_1$ to have dimension 
1/mass, whereas the remaining variables $\tau_i$ for $i\geq2$ should be dimensionless. 
Our ansatz \eqref{eq:measure:NLO} for the one-emission measurement function can then be 
generalised to
\begin{align}
\label{eq:M1:multidiff}
\mathcal{M}_1(\tau_1,\tau_2,\ldots; k) = \exp\big(-\tau_1\, k_{T}\, y_k^{n/2}\, 
f(y_k,t_k;\tau_2,\ldots)\,\big)\,,
\end{align}
and similarly for the two-emission functions. As long as the RGE \eqref{eq:RGE:SCET-1} 
only depends on the first Laplace variable through logarithms $L_1=\ln(\mu\taubar_1)$, 
the results from Section~\ref{sec:SCET-1} for the soft anomalous dimension can equally 
be applied for multi-differential soft functions. The same is true for the expressions 
from Section~\ref{sec:SCET-2} for the collinear anomaly exponent as long as the expansion 
in \eqref{eq:d1d2} only depends on $L_1$. 

As an example of a double-differential soft function, we consider the one for exclusive 
Drell-Yan production from~\cite{Li:2011zp}. Due to rescaling invariance of the Wilson lines, 
the position-space soft function can only depend on $\tau_1 = \sqrt{x_+ x_-}$ and 
$\tau_2 = \sqrt{x_T^2/x_+x_-}$, which play the role of a dimensionful and a dimensionless 
Laplace (or Fourier) variable. After rescaling $\tau_1\to \frac{i}{2}\tau_1$, the soft 
function is then specified by $n=-1$, $f(y_k,t_k;\tau_2) = 1 + y_k - 2\sqrt{y_k}(1-2t_k)\tau_2$ 
and
\begin{align}
F(a,b,y,t_k,t_l,t_{kl};\tau_2) = 
1 + y - 2 \;\sqrt{\frac{a y}{(1+a b)(a+b)}} \;
\Big(b (1-2t_k) + 1-2t_l\Big)\tau_2\,.
\end{align}
As the underlying RGE only depends on logarithms of $\tau_1$~\cite{Li:2011zp}, we can apply 
the formulae from Section~\ref{sec:SCET-1} to compute the soft anomalous dimension. This yields 
$\gamma^{S}_0 =0$, $\gamma^{n_f}_1 =224/27 - 4/9\,\pi^2$ and 
$\gamma^{C_A}_1 =-808/27 + 11/9\,\pi^2 + 28\zeta_3$, along with $\gamma^{C_F}_1=0$ since 
the soft function is consistent with NAE in position space. These findings are in agreement 
with~\cite{Li:2011zp}. 

We next consider the double-differential hemisphere soft function that was computed to 
NNLO in~\cite{Kelley:2011ng,Hornig:2011iu}. In this case, we take two Laplace transformations 
with respect to the hemisphere masses $M_L$ and $M_R$ and denote the respective Laplace 
variables by $\tau_L$ and $\tau_R$. We may then choose~$\tau_1=\sqrt{\tau_L\tau_R}$ and 
$\tau_2=(\tau_L+\tau_R)/\sqrt{\tau_L\tau_R}$, which is convenient since these variables
respect the $\tau_L\leftrightarrow \tau_R$ symmetry of the soft function. More importantly, 
the RGE then again only depends on logarithms of $\tau_1$, such that the formulae 
from Section~\ref{sec:SCET-1} can be used to compute the soft anomalous dimension. 
Without going into further details here, we find the same result as in the previous example, 
which is in line with the findings of~\cite{Kelley:2011ng,Hornig:2011iu}.\\[-0.5em] 

\noindent
\textbf{N-jet soft functions:}
The computation of soft functions that involve Wilson lines in more than two light-like 
directions is clearly more complicated and beyond the scope of the present paper. 
Still, it has been argued in~\cite{Becher:2015lmy} that the anomalous dimension of an 
$N$-jet soft function can be reconstructed from the information on dijet soft functions. 
The strategy of~\cite{Becher:2015lmy} relies on the fact that the two-loop hard anomalous 
dimensions are known for arbitrary processes~\cite{Becher:2009qa} and that cross sections 
are invariant under a variation of the factorisation scale. 

The authors of~\cite{Becher:2015lmy} illustrate this method with the hadronic event-shape 
variable transverse thrust. In the dijet limit, the transverse thrust distribution satisfies 
a hard-beam-jet-soft factorisation theorem that contains a soft function which depends on two 
incoming and two outgoing light-like directions~\cite{Becher:2015gsa}. By considering simpler 
toy processes, in which all but two of the hard QCD partons are replaced by leptons, the required 
jet and beam anomalous dimensions can then be extracted from known results of hard and soft 
(dijet) anomalous dimensions. This information is then used in a second step to determine the 
soft anomalous dimension of the $N$-jet observable.

The toy processes that are needed to determine the two-loop soft anomalous dimension for 
transverse thrust are $e^+ e^-\to q\bar q$ and $q\bar q \to e^+ e^-$ scattering. In the first 
case, the soft function falls into the pattern defined in Section~\ref{sec:measurement} 
with $n=1$, $f(0,t_k) =16t_k\bar t_k$ and
\begin{align}
F_A(a,b,0,t_k,t_l,t_{kl}) &= \frac{16(a t_l \bar t_l + b t_k \bar t_k)}{a+b}\,,
\nonumber\\
F_B (a,b,0,t_k,t_l,t_{kl}) &= \frac{16(t_l\bar t_l+abt_k\bar t_k)}{1+ab}\,.
\end{align}
The integral representations for the calculation of the soft anomalous dimension from 
Section~\ref{sec:SCET-1} can then be evaluated numerically, giving $\gamma^{S}_0 =0$, 
$\gamma^{n_f}_1 =19.3954(5)$ and $\gamma^{C_A}_1 =-158.276(5)$, 
as well as $\gamma^{C_F}_1 =0$ since the soft function is consistent with NAE.
The two-loop soft anomalous dimension for this (toy) observable was previously extracted 
via an {\tt EVENT2} fit in~\cite{Becher:2015gsa} with considerably larger uncertainties 
($\gamma^{n_f}_1 =18^{+2}_{-3}$, $\gamma^{C_A}_1 =-148^{+20}_{-30}$).

The soft function for the second toy process turns out to be a SCET-2 observable with
$n=0$, $f(y_k,t_k) =2 c_0 (1-|1-2t_k|)$ and
\begin{align}
F(a,b,y,t_k,t_l,t_{kl}) &= 2 c_0\;
\sqrt{\frac{a}{(1+a b)(a+b)}}\;
\big(b (1 - |1 - 2 t_k|) + 1 - |1 - 2 t_l|\big)\,,
\end{align}
where $c_0=e^{4G/\pi}$ involves Catalan's constant $G\simeq 0.915966$. In this case we
obtain $d_1 =0$, $d^{\,n_f}_2 =-37.1743(5)$ and $d_2^{\,C_A} =208.098(3)$, 
whereas again $d_2^{\,C_F} =0$ because of NAE. These numbers can be compared to the 
calculation from~\cite{Becher:2015lmy}, which quotes 
$d^{\,n_f}_2 =-37.191(6)$ and $d_2^{\,C_A} =208.0(1)$.

As the strategy proposed in~\cite{Becher:2015lmy} is general, we conclude that our results for 
dijet soft functions can be used indirectly to determine soft anomalous dimensions for processes 
with more than two light-like directions.

\section{Conclusions}
\label{sec:conclusions}

We have developed a novel formalism for the calculation of two-loop soft anomalous dimensions
that is relevant for processes with two hard, massless and colour-charged partons. As long as 
the corresponding soft function falls into the pattern defined in Section~\ref{sec:measurement}, 
the integral representations for the soft anomalous dimensions can easily be evaluated 
numerically, without having to perform an explicit two-loop calculation anymore. Our approach 
is sufficiently general to treat observables that are defined in SCET-1 and SCET-2, and we
clarified the relation between the respective soft anomalous dimension and the collinear 
anomaly exponent.

By considering various examples, we illustrated that our setup can be applied to a large
variety of dijet soft functions. In particular, we computed the two-loop soft anomalous 
dimension of the $e^+ e^-$ angularity event shape and the soft-drop jet-grooming algorithm 
for the first time. Our results allow one to extend existing resummations for these observables 
to NNLL accuracy. In Section~\ref{sec:generalisation} we have furthermore shown that our
formalism can be generalised to soft functions which a priori do not belong to the class
defined in Section~\ref{sec:measurement}. This includes, in particular, jet-veto observables 
and soft functions that are relevant for processes with more than two jet directions.

We believe that our results will help facilitate precision resummations in 
both QCD and SCET in the future, and that they may be particularly useful for developing an 
automated resummation code. For convenience of the user, we plan to implement the integral 
representations from this paper in the forthcoming {\tt SoftSERVE} 
distribution~\cite{Bell:2018oqa}.\\[-0.5em]

{\em Acknowledgments:\/}
We are grateful to Thomas Becher and Bahman Dehnadi for helpful discussions and comments on the 
manuscript. G.B.~is supported by the Deutsche Forschungsgemeinschaft (DFG) within Research Unit 
FOR 1873. R.R.~is supported by the Swiss National Science Foundation (SNF) under grant 
CRSII2-160814. J.T.~acknowledges research and travel funds from DESY. G.B. and R.R. thank the 
Munich Institute for Astro- and Particle Physics (MIAPP) of the DFG cluster of excellence 
''Origin and Structure of the Universe'' for hospitality and support.

\begin{appendix}

\section{Details of the $C_F^2$ contribution}
\label{app:CF2:details}

For the uncorrelated emission contribution, we find that the pole terms of the bare soft 
function only cancel as predicted by the RGE if the following constraint is satisfied,
\begin{align}
& 
\frac{8}{\pi}\,
\int_0^1 \!dt_l \;\,
\frac{\ln^2 f(0,t_l)}{\altsqrt{4t_l\bar t_l}}
-\frac{16}{\pi^2}\,
\int_0^1 \!dt_k \;\,
\frac{\ln f(0,t_k)}{\altsqrt{4t_k\bar t_k}}\;
\int_0^1 \!dt_l \;\,
\frac{\ln f(0,t_l)}{\altsqrt{4t_l\bar t_l}}
\nonumber\\[0.2em]
&\qquad
-\frac{4}{\pi^2}
\int_0^1 \!db \int_0^1 \!dt_l \int_0^1 \!dt_{kl} \;
\frac{1}{\altsqrt{16t_l\bar t_l t_{kl}\bar t_{kl}}}\; \,
\frac{\mathcal{G}_1(0,b,t_l,t_{kl})}{b_+}
\,=\,0\,,
\label{eq:CF2:constaint}
\end{align}
where the explicit form of the function $\mathcal{G}_{1}(y,b,t_l,t_{kl})$ was given in 
\eqref{eq:G1G2}. We checked that this constraint is fulfilled for all soft functions 
we considered explicitly in this work, but we cannot prove that it holds in the 
general case.

Similarly, we find an additional contribution to the two-loop soft anomalous dimension
$\gamma^{C_F}_1$ and the two-loop anomaly exponent $d^{\,C_F}_2$, which vanishes for all 
examples we considered, and which we conjecture to be zero in general. This contribution 
reads
\begin{align}
\Delta \gamma^{C_F}_1 &=
\frac{64}{n} \,\bigg\{ 
\frac{4}{\pi}\,
\int_0^1 \!dt_l \;\,
\frac{\ln^3 f(0,t_l)}{\altsqrt{4t_l\bar t_l}}
-\frac{2}{\pi}\,
\int_0^1 \!dt_l \;\,
\frac{\ln(16t_l\bar t_l)}{\altsqrt{4t_l\bar t_l}}\;\ln^2 f(0,t_l)
\nonumber\\[0.2em]
&\qquad\quad
-\frac{8}{\pi^2}\,
\int_0^1 \!dt_k \;\,
\frac{\ln f(0,t_k)}{\altsqrt{4t_k\bar t_k}}\;
\int_0^1 \!dt_l \;\,
\frac{\ln f(0,t_l)}{\altsqrt{4t_l\bar t_l}}\;
\ln \frac{f(0,t_l)}{16t_l\bar t_l}
\nonumber\\[0.2em]
&\qquad\quad
+\frac{1}{\pi^2}
\int_0^1 \!db \int_0^1 \!dt_l \int_0^1 \!dt_{kl} \;
\frac{1}{\altsqrt{16t_l\bar t_l t_{kl}\bar t_{kl}}}\; \,
\bigg[\frac{1}{b} \,\ln \frac{256\,t_l\bar t_l t_{kl}\bar t_{kl}\,b^2}{(1+b)^4} \bigg]_+\;\mathcal{G}_1(0,b,t_l,t_{kl})
\nonumber\\[0.2em]
&\qquad\quad
-\frac{2}{\pi^2}
\int_0^1 \!db \int_0^1 \!dt_l \int_0^1 \!dt_{kl} \;
\frac{1}{\altsqrt{16t_l\bar t_l t_{kl}\bar t_{kl}}}\; \,
\frac{\mathcal{G}_3(b,t_l,t_{kl})}{b_+}
\nonumber\\[0.2em]
&\qquad\quad
+\frac{2}{\pi^2}
\int_0^1 \!db \int_0^1 \!dt_l \int_0^1 \!dt_{kl} \int_0^1 \!ds \;
\frac{1}{\altsqrt{16t_l\bar t_l t_{kl}\bar t_{kl}}}\; \,
\frac{1}{b} \bigg[\frac{1}{s\sqrt{1-s^2}} \bigg]_+ \;\mathcal{G}_4(b,t_l,t_{kl},s)\bigg\}
\label{eq:CF2:deltagamma1}
\end{align}
for the soft anomalous dimension, and
\begin{align}
\Delta d^{\,C_F}_2 &=
64 \,\bigg\{ 
-\frac{4}{\pi}\,
\int_0^1 \!dt_l \;\,
\frac{\ln^3 f(0,t_l)}{\altsqrt{4t_l\bar t_l}}
+\frac{8}{\pi^2}\,
\int_0^1 \!dt_k \;\,
\frac{\ln f(0,t_k)}{\altsqrt{4t_k\bar t_k}}\;
\int_0^1 \!dt_l \;\,
\frac{\ln^2 f(0,t_l)}{\altsqrt{4t_l\bar t_l}}\;
\nonumber\\[0.2em]
&\qquad\quad
-\frac{1}{\pi^2}
\int_0^1 \!db \int_0^1 \!dt_l \int_0^1 \!dt_{kl} \;
\frac{1}{\altsqrt{16t_l\bar t_l t_{kl}\bar t_{kl}}}\; \,
\bigg[\frac{1}{b} \,\ln \frac{b^2}{(1+b)^4} \bigg]_+\;\mathcal{G}_1(0,b,t_l,t_{kl})
\nonumber\\[0.2em]
&\qquad\quad
+\frac{2}{\pi^2}
\int_0^1 \!db \int_0^1 \!dt_l \int_0^1 \!dt_{kl} \;
\frac{1}{\altsqrt{16t_l\bar t_l t_{kl}\bar t_{kl}}}\; \,
\frac{\mathcal{G}_3(b,t_l,t_{kl})}{b_+}\bigg\}
\label{eq:CF2:deltad2}
\end{align}
for the collinear anomaly exponent. Here
\begin{align}
\mathcal{G}_{3}(b,t_l,t_{kl}) &=
\ln^2 G_{A_1}(0,0,b,t_k^+,t_l,t_{kl})
+ \ln^2 G_{A_2}(0,0,b,t_k^+,t_l,t_{kl})
\nonumber\\
&\quad
+ \ln^2 G_{B_1}(0,0,b,t_k^+,t_l,t_{kl})
+ \ln^2 G_{B_2}(0,0,b,t_k^+,t_l,t_{kl})
+ (t_k^+\to t_k^-)\,,
\nonumber\\
\mathcal{G}_{4}(b,t_l,t_{kl},s) &=
\ln G_{A_1}(0,0,b,t_k^{\oplus},t_l,t_{kl})
+ \ln G_{A_2}(0,0,b,t_k^{\oplus},t_l,t_{kl})
\nonumber\\
&\quad
+ \ln G_{B_1}(0,0,b,t_k^{\oplus},t_l,t_{kl})
+ \ln G_{B_2}(0,0,b,t_k^{\oplus},t_l,t_{kl})
+ (t_k^{\oplus}\to t_k^{\ominus})\,,
\end{align} 
and
\begin{align}
t_k^{\oplus} &= t_l + t_{kl} - 2 t_l t_{kl} + 2 \sqrt{t_l\bar t_l t_{kl}\bar t_{kl}(1-s^2)}\,,
\nonumber\\
t_k^{\ominus} &= t_l + t_{kl} - 2 t_l t_{kl} - 2 \sqrt{t_l\bar t_l t_{kl}\bar t_{kl}(1-s^2)}\,.
\end{align}

For the cumulant soft functions $\widehat{S}(\omega,\mu)$ discussed in 
Section~\ref{sec:generalisation}, the above relations are slightly modified. In particular, 
we find an additional term $-2\pi^2/3$ on the left hand side of equation 
\eqref{eq:CF2:constaint} and, similarly, the corresponding relations to 
\eqref{eq:CF2:deltagamma1} and \eqref{eq:CF2:deltad2} for cumulant soft functions become
\begin{align}
\Delta \widehat{\gamma}^{C_F}_1 &= \Delta \gamma^{C_F}_1 
-\frac{256\zeta_3}{n} 
+ \frac{16\pi^2}{3n}  \,\frac{\gamma^{S}_0}{C_F}\,,
\nonumber\\
\Delta \widehat{d}^{\,C_F}_2 &= \Delta d^{\,C_F}_2 
+256\zeta_3 
+ \frac{16\pi^2}{3} \,\frac{d_1}{C_F}\,,
\end{align}
which we again conjecture to vanish for all observables.

\section{Details of cumulant soft functions}
\label{app:cumulant:details}

In this appendix we list the explicit expressions for the measurement functions of the 
three cumulant soft functions discussed in Section \ref{sec:generalisation}. These are 
required to compute the soft anomalous dimensions and the collinear anomaly exponents from
Figures~\ref{fig:jetvetoes} and \ref{fig:jetgrooming}.  
\\[-0.5em]

\textbf{Rapidity-dependent jet vetoes:}
As the four jet-veto observables from~\cite{Gangal:2014qda} have the same soft anomalous 
dimension, we focus here on the C-parameter jet veto in the hadronic center-of-mass frame, 
$\mathcal{T}_{C\rm cm}$, for concreteness. The corresponding soft function is specified by 
$n=1$, 
$f(y_k,t_k) =1/(1+y_k)$ and
\begin{align}
F_A(a,b,0,t_k,t_l,t_{kl}) &= \theta(\Delta_F - R) \;\frac{\text{max}(a, b)}{a + b} +
\theta(R - \Delta_F)\,,
\nonumber\\
F_B(a,b,0,t_k,t_l,t_{kl}) &=\theta(\Delta_F - R) \;\frac{\text{max}(1,a b)}{1 +a b} +
\theta(R - \Delta_F)\,,
\end{align}
where $R$ is the jet radius and $\Delta_F= \sqrt{\ln^2 a+\arccos^2(1-2t_{kl})}$ represents 
the clustering condition in the parametrisation \eqref{eq:param:NNLO:corr}. For uncorrelated 
emissions, we find
\begin{align}
G_{i}(y,0,b,t_k,t_l,t_{kl}) &= \frac{1}{(1 + b)(1 + y)}\,,
\hspace{37.7mm} (i=A_1,A_2,B_1,B_2)
\nonumber\\
G_{i}(0,r,b,t_k,t_l,t_{kl}) &= \theta(\Delta_G - R) \;\frac{1}{1 + b} +
\theta(R - \Delta_G)\,,
\qquad (i=A_1,A_2)
\nonumber\\
G_{i}(0,r,b,t_k,t_l,t_{kl}) &= \frac{1}{1 + b}\,,
\hspace{53.7mm} (i=B_1,B_2)
\end{align}
where $\Delta_G= \sqrt{\frac14 \ln^2 r+\arccos^2(1-2t_{kl})}$ is the analogous clustering 
constraint in the parametrisation \eqref{eq:param:NNLO:unc}.\\[-0.5em]

\textbf{Standard jet veto:}
The $p_T$ jet veto turns out to be a SCET-2 observable that depends on the same clustering 
conditions in terms of $\Delta_F$ and $\Delta_G$ as in the previous example. The corresponding 
soft function satisfies $n=0$, $f(y_k,t_k) =1$ and
\begin{align}
F_i(a,b,0,t_k,t_l,t_{kl}) &= \sqrt\frac{a}{(a + b)(1 + a b)}
\Big( \theta(\Delta_F - R) + \theta(R - \Delta_F) \;
\sqrt{1 + b^2 + 2 b (1 - 2 t_{kl})} \Big)
\end{align}
for both regions $i=A,B$. For uncorrelated emissions, we now obtain
\begin{align}
G_{i}(y,0,b,t_k,t_l,t_{kl}) &= \frac{1}{1+b}\,,
\hspace{64.0mm} (i=A_1,A_2,B_1,B_2)
\nonumber\\
G_{i}(0,r,b,t_k,t_l,t_{kl}) &= \theta(\Delta_G - R) \;\frac{1}{1 + b} 
\nonumber\\
&\quad+
\theta(R - \Delta_G)\;\frac{\sqrt{1 + b^2 + 2 b (1 - 2 t_{kl})}}{1 + b}\,,
\qquad (i=A_1,A_2)
\nonumber\\
G_{i}(0,r,b,t_k,t_l,t_{kl}) &= \frac{1}{1 + b}\,.
\hspace{63.9mm} (i=B_1,B_2)
\end{align}\\[-1.5em]

\textbf{Jet grooming:}
The soft function for the soft-drop jet grooming algorithm is characterised by $n= -1-\beta$, 
$f(y_k,t_k) = (1 + y_k)^{1 + \beta/2}$ (for $0\leq y_k\leq1$) and
\begin{align}
F_A(a,b,0,t_k,t_l,t_{kl}) &= 1+ 
\theta(1+4a t_{kl}-2a) \,\left(a^{-\beta/2} (a + b)^{\beta/2} (1 + a b)^{-1 - \beta/2}-1\right)\,,
\nonumber\\
F_B(a,b,0,t_k,t_l,t_{kl}) &=1+ 
\theta(1+4a t_{kl}-2a) 
\nonumber\\
&\qquad
\Big\{ \theta(b-a^{1+\beta}) \, b \,a^{-\beta/2} (a + b)^{-1-\beta/2} (1 + a b)^{\beta/2}
\nonumber\\
&\qquad\quad
+\theta(a^{1+\beta}-b)  \,a^{1+\beta/2} (a + b)^{-1-\beta/2} (1 + a b)^{\beta/2} -1 \Big\}\,.
\end{align}
The measurement function for the uncorrelated emission contribution is in this case given by
\begin{align}
G_{i}(y,0,b,t_k,t_l,t_{kl}) &= \frac{(1 + y)^{1 + \beta}}{1 + b}\,,
\hspace{85.1mm} (i=A_1,B_1)
\nonumber\\
G_{i}(y,0,b,t_k,t_l,t_{kl}) &= \theta_1 \,
\frac{(1 + y)^{1 + \beta/2}}{1 + b} 
+ (1-\theta_1) \,\frac{b (1 + y)^{1 + \beta}}{1 + b}\,,
\hspace{37mm} (i=A_2,B_2)
\nonumber\\
G_{A_1}(0,r,b,t_k,t_l,t_{kl}) &= 
\theta_2 \theta_3 \,
\frac{(b + r^{1 + \beta/2})^{-\beta/2} (b + r^{\beta/2})^{1 + \beta/2}}{1 + b}
+ (1 -  \theta_2 \theta_3)\,\frac{1}{1 + b}\,,			
\nonumber\\
G_{A_2}(0,r,b,t_k,t_l,t_{kl}) &= 
\theta_2 \theta_3 \,
\frac{(b\, r^{1 + \beta/2}+1)^{-\beta/2} (b\, r^{\beta/2}+1)^{1 + \beta/2}}{1 + b}
+ (1 -  \theta_2 \theta_3)\,\frac{1}{1 + b}\,,			
\nonumber\\
G_{i}(0,r,b,t_k,t_l,t_{kl}) &= \frac{1}{1 + b}\,,
\hspace{94.4mm} (i=B_1,B_2)
\end{align}
where $\theta_1=\theta\big(1 - b (1 + y)^{\beta/2}\big)$,
$\theta_2=\theta\big(2 (1 - 2 t_{kl}) - \sqrt{r}\big)$ and
$\theta_3=\theta\big(2 \sqrt{r} (1 - 2 t_{kl})-1\big)$.

\end{appendix}

\end{document}